\documentclass[journal,onecolumn]{IEEEtran}
\usepackage[utf8]{inputenc}
\usepackage{units}
\usepackage{placeins}
\usepackage{flafter}
\usepackage{cite}
\usepackage{amssymb}
\usepackage[cmex10]{amsmath}
\DeclareMathOperator*{\argmin}{\arg\!\min}
\DeclareMathOperator*{\argmax}{\arg\!\max}
\usepackage{algpseudocode}
\usepackage{mdwmath}
\usepackage{mdwtab}
\newcolumntype{C}[1]{>{\centering \hspace{0pt}}p{#1}}
\usepackage[caption=false,font=footnotesize]{subfig}
\usepackage{fixltx2e}
\usepackage{hyperref}
\newlength\figureheight
\newlength\figurewidth
\usepackage{tikzexternal}
\renewcommand*{\Re}{\operatorname{Re}}
\renewcommand*{\Im}{\operatorname{Im}}

\IEEEpubid{Published in: IEEE Transactions on Signal Processing, vol. 63, no. 24, pp. 6606 - 6615, Dec. 2015, DOI: \href{http://dx.doi.org/10.1109/TSP.2015.2477052}{10.1109/TSP.2015.2477052} \copyright~2015 IEEE}

\begin{document}

\title{Coherence Optimization and\\ Best Complex Antipodal Spherical Codes}
\author{Henning Zörlein,~\IEEEmembership{Student Member,~IEEE}, and Martin Bossert,~\IEEEmembership{Fellow,~IEEE}
\thanks{This work was supported by the German research council Deutsche Forschungsgemeinschaft (DFG) under Grant Bo~867/27-1.
The authors are with the Institute of Communications Engineering at Ulm University in Ulm, Germany.\newline
E-Mail: \{henning.zoerlein, martin.bossert\}@uni-ulm.de
}
}

\maketitle

\begin{abstract}
Vector sets with optimal coherence according to the Welch bound cannot exist for all pairs of dimension and cardinality.
If such an optimal vector set exists, it is an equiangular tight frame and represents the solution to a Grassmannian line packing problem.
Best Complex Antipodal Spherical Codes (BCASCs) are the best vector sets with respect to the coherence.
By extending methods used to find best spherical codes in the real-valued Euclidean space, the proposed approach aims to find BCASCs, and thereby, a complex-valued vector set with minimal coherence.
There are many applications demanding vector sets with low coherence.
Examples are not limited to several techniques in wireless communication or to the field of compressed sensing.
Within this contribution, existing analytical and numerical approaches for coherence optimization of complex-valued vector spaces are summarized and compared to the proposed approach.
The numerically obtained coherence values improve previously reported results.
The drawback of increased computational effort is addressed and a faster approximation is proposed which may be an alternative for time critical cases.
\end{abstract}


\IEEEpeerreviewmaketitle
\section{Introduction}
\IEEEPARstart{T}{he} coherence of vector sets is an important and a limiting factor in many applications like Multiple-Input Multiple-Output~(MIMO) and Code Division Multiple Access~(CDMA) wireless systems, non-orthogonal multi-pulse modulation and Compressed Sensing~(CS)~\cite{HeathStrohmerPaulraj2006,hochwald_systematic_2000,gohary_noncoherent_2009,schober_geodesical_2009,mccloud_modulation_2002,FoucartRauhut2013,LazichZoerleinBossertSCC2013,li_deterministic_2014}.
Therefore, research in coherence optimization is of natural interest.

The problem of coherence optimization is related to several other well known optimization problems, e.g. Grassmannian line packing, sphere packing, minimum distance optimization and frame design, consequently, the field has already a large history~\cite{LoveGrassmannians2003,ConwayHardinSloane1996,fickus_steiner_2012}. 
Often, only the case of real vector spaces is considered.
In this paper, the more general complex case shall be of central interest.
Only a few, very specific analytical (nearly) optimal solutions are known.
However, several different numerical approaches have been proposed in the last decades.
Since the optimization problem is very challenging, there is still ongoing research.
Therefore, we propose a concept for coherence minimization which is based on distance optimization of Complex Antipodal Spherical Codes (CASCs), since it is shown that Best Complex Antipodal Spherical Codes (BCASCs) result in vector sets of minimal coherence.
The performance of the proposed scheme is numerically evaluated and compared to existing alternatives for coherence minimization.

The remainder of this paper is structured as follows:
In Section~\ref{sec:Math}, the optimization problem and the necessary notation is defined and connected to minimal distance maximization and Grassmannian line packing.
The concept of BCASCs is introduced in Section~\ref{sec:BCASC} and used for coherence optimization.
The proposed approach is discussed in Section~\ref{sec:discussion}.
The success of the optimization is numerically evaluated in Section~\ref{sec:NumSim}.

\section{Notation and Prerequisites}
\label{sec:Math}
Matrices are denoted by bold capital letters, e.g., $\boldsymbol{A}\in\mathbb{C}^{N\times M}$, and vectors by bold lower-case letters, e.g., $\boldsymbol{s}\in\mathbb{C}^N$ throughout the paper.
Scalar values are represented by non-bold letters, e.g., $\alpha\in\mathbb{R}$.
In the following, if not stated otherwise, the $N$-dimensional complex vector space $\mathbb{C}^{N}$ is considered.
Consequently, the inner product between two vectors $\boldsymbol{s}_1,\boldsymbol{s}_2\in\mathbb{C}^N$ is defined as $\left<\boldsymbol{s}_1,\boldsymbol{s}_2\right>=\boldsymbol{s}_1^{\mathrm{H}}\boldsymbol{s}_2$ and the norm as ${\|\boldsymbol{s}_1\|}=\sqrt{\left<\boldsymbol{s}_1,\boldsymbol{s}_1\right>}\in\mathbb{R}$.
A complex number $s=s^{\operatorname{R}}+\mathrm{i}\cdot s^{\operatorname{I}}\in\mathbb{C}$ consists of a real part $s^{\operatorname{R}}=\Re(s)$ and an imaginary part $s^{\operatorname{I}}=\Im(s)$.

\subsection{Coherence, the Welch Bound and Beyond}
A vector set $\mathcal{A}$ can be represented by a matrix $\boldsymbol{A}\in\mathbb{C}^{N\times M}$, where the $M$ columns of dimension $N$ correspond to the vectors of $\mathcal{A}$.
The coherence of $\boldsymbol{A}$ is defined as
\begin{equation}
\label{eq:coherence}
\mu(\boldsymbol{A}) = \max \limits_{i\neq j} \frac{|\langle\boldsymbol{a}_i,\boldsymbol{a}_j\rangle|}{{{\|\boldsymbol{a}_i\|}}{{\|\boldsymbol{a}_j\|}}},
\end{equation}
where $\boldsymbol{a}_i$ denotes the $i$-th column.
For the case of column-normalized matrices (e.g., those corresponding to spherical codes), the normalization is often omitted in the definition: $\mu(\boldsymbol{A}) = \max \limits_{i\neq j} |\langle\boldsymbol{a}_i,\boldsymbol{a}_j\rangle|$.

In the following, the non-trivial case of $M>N$ is considered, where the value of the coherence $\mu(\boldsymbol{A})$ is lower bounded by the so called Welch bound:
\begin{equation}
\label{eq:welch}
\mu(\boldsymbol{A})\geq\sqrt{\frac{M-N}{N(M-1)}}
\end{equation}
This bound has been actually stated first for the real-valued case $\boldsymbol{A}\in\mathbb{R}^{N\times M}$ by Rankin~\cite{Rankin1956} and for the complex-valued case by Welch~\cite{Welch1974}.
Therefore, it is sometimes also called Rankin or simplex bound~\cite{ConwayHardinSloane1996}.

It is desirable to obtain a set of vectors $\mathcal{A}$ which achieves equality in~\eqref{eq:welch}.
However, two different criteria for determining whether a set satisfies the Welch bound with equality have been established in literature~\cite{sarwate_meeting_1999}.
Massey et al.~\cite{massey_welchs_1993} consider the Root Mean Square (RMS) magnitude of the inner product and denote vector sets as Welch Bound Equality (WBE) sequences for which 
\begin{equation}
\label{eq:WBE_Def}
\sqrt{\frac{1}{M(M-1)}\sum_{k=1}^{M}\sum_{\substack{l=1\\ l\neq k}}^{M}|\langle\boldsymbol{a}_k,\boldsymbol{a}_l\rangle|^2}=\sqrt{\frac{M-N}{N(M-1)}}
\end{equation} 
is fulfilled.
However, the maximal inner product is considered in the majority of publications as it is implied by equality of~\eqref{eq:coherence} and~\eqref{eq:welch}.
The corresponding sets are denoted as Maximum Welch Bound Equality (MWBE) sequences by~\cite{sarwate_meeting_1999} and form a subclass of WBE sequences.
For the remainder of this paper, only the maximal inner product, and therefore, MWBE sequences will be considered.

According to~\cite{DelsarteGoethalsSeidel1975}, MWBE sequences can only exist if
\begin{align}
\nonumber M&\leq\frac{N(N+1)}{2}, &&\text{ for }\boldsymbol{A}\in\mathbb{R}^{N\times M}\\
\label{eq:WelchLimits}M&\leq N^2, &&\text{ for }\boldsymbol{A}\in\mathbb{C}^{N\times M}.
\end{align}
It should be noted that these are just necessary conditions since cases are shown in~\cite{ConwayHardinSloane1996} for which equality in \eqref{eq:welch} cannot be achieved.

Settings with large $M$, in which the Welch bound cannot be met with equality according to \eqref{eq:WelchLimits}, are considered  by several bounds.
For example, the orthoplex bound is given in~\cite{ConwayHardinSloane1996} for the real-valued case, and has been subsequently extended to the complex-valued case in~\cite{henkel_sphere-packing_2005,pitaval_low_2011}:
\begin{align}
\label{eq:orthoplex}
\mu(\boldsymbol{A})&\geq\sqrt{\frac{1}{N}}
\end{align}
The orthoplex bound is only achievable if
\begin{align}
\nonumber \frac{N(N+1)}{2}&<M\leq (N-1)(N+2), &&\text{ for }\boldsymbol{A}\in\mathbb{R}^{N\times M}\\
 N^2&<M\leq 2(N^2-1), &&\text{ for }\boldsymbol{A}\in\mathbb{C}^{N\times M}.
\end{align}
Levenshtein developed another bound for the described case of too many vectors~\cite{levenshtein_bounds_1978,levenshtein_bounds_1983,ding_signal_2007}:
\begin{align}
\nonumber \mu(\boldsymbol{A})&\geq\sqrt{\frac{3M-N^2-2N}{(N+2)(M-N)}}, &&\text{ for }\boldsymbol{A}\in\mathbb{R}^{N\times M}\\
\label{eq:Levenshtein}\mu(\boldsymbol{A})&\geq\sqrt{\frac{2M-N^2-N}{(N+1)(M-N)}}, &&\text{ for }\boldsymbol{A}\in\mathbb{C}^{N\times M}
\end{align}
And a few decades later, a further bound for this case has been derived in~\cite{mukkavilli_beamforming_2003,xia_achieving_2005}:
\begin{equation}
\label{eq:BoundMukkavilli}
\mu(\boldsymbol{A})\geq1-2M^{-\frac{1}{N-1}}
\end{equation}

Taking the maximum over all mentioned bounds, within their corresponding regimes, results in the following composite lower bound for the complex case:
\begin{equation}
\label{eq:LowerBoundWhereWelchBoundFails}
\textstyle
\mu(\boldsymbol{A})
\geq 
\left\{
\begin{array}{l}
\text{for }\ M\leq N^2:\\
\sqrt{\frac{M-N}{N(M-1)}}\\
\\
\text{for }\ N^2<M\leq 2(N^2-1):\\
\max\left\{ \sqrt{\frac{1}{N}},\sqrt{\frac{2M-N^2-N}{(N+1)(M-N)}}, 1-2M^{-\frac{1}{N-1}}\right\}\hfill\\
\\
\text{for }\ 2(N^2-1)<M:\\
\max\left\{ \sqrt{\frac{2M-N^2-N}{(N+1)(M-N)}}, 1-2M^{-\frac{1}{N-1}}\right\}\hfill\\
\end{array}\right.
\end{equation}

\subsection{Frame Theory} 
The concept of frame theory has been introduced in 1952~\cite{duffin_class_1952}.
In the following, only finite frames are considered.
Because of $M>N$, a frame can be interpreted as an overcomplete basis.
A set $\mathcal{A}=\left\{\boldsymbol{a}_i\right\}_1^M$ of $M$ vectors spanning $\mathbb{C}^N$ is denoted as frame if there exist two constants $0< A\leq B <\infty$, such that for all $\boldsymbol{x}\in\mathbb{C}^N$
\begin{equation}
A\|\boldsymbol{x}\|^2\leq\sum\limits_{i=1}^{M}|\langle\boldsymbol{a}_i,\boldsymbol{x}\rangle|^2\leq B\|\boldsymbol{x}\|^2,
\end{equation}
where $A,B\in\mathbb{R}$ are the so called frame bounds.
If $A=B$, $\mathcal{A}$ is denoted as $A$-tight frame.
Consequently, the rows of the corresponding matrix $\boldsymbol{A}$ are of equal-norm and orthogonal to each other.
If $\|\boldsymbol{a}_i\|=1$ for $i=1\hdots M$, $\mathcal{A}$ is called a unit norm frame.
Thus, a Unit Norm Tight Frame (UNTF) has necessarily a frame bound $A=M/N$ also known as the frame redundancy.
An Equiangular Tight Frame (ETF) has the additional property $|\langle\boldsymbol{a}_i,\boldsymbol{a}_j\rangle|=\mu\ \forall\ i\neq j $.
By this definition, an ETF consists out of MWBE sequences.
A detailed introduction to frame theory can be found in~\cite{kovacevic_life01_2007,kovacevic_life02_2007,casazza_introduction_2013}.
In~\cite{strohmer_grassmannian_2003}, frames minimizing \eqref{eq:coherence} are defined as Grassmannian frames.
By definition, an ETF is an optimal Grassmannian frame.

For a given full-rank matrix $\boldsymbol{A}$, the closest $B$-tight frame in Frobenius norm can be calculated by $B(\boldsymbol{A}\boldsymbol{A}^{\mathrm{H}})^{-\nicefrac{1}{2}}\boldsymbol{A}$~\cite{tropp_designing_2005}.
This can be used to obtain a $B$-tight frame which is close to a (numerically obtained) non-optimal Grassmannian frame.

In~\cite{benedetto_finite_2003}, the frame potential 
\begin{equation}
\text{FP}(\left\{\boldsymbol{a}_i\right\}_1^M)=\sum\limits_{k=1}^{M}\sum\limits_{l=1}^{M}|\langle\boldsymbol{a}_k,\boldsymbol{a}_l\rangle|^2
\end{equation}
is derived from a frame force, which is notably different from the forces subsequently defined in this paper.
Minimizing the frame potential results in UNTF~\cite{benedetto_finite_2003}.
For these frames, $\text{FP}(\left\{\boldsymbol{a}_i\right\}_1^M)$ equals \eqref{eq:WBE_Def} up to a constant factor, and thus, a UNTF consists out of WBE sequences.

\subsection{Grassmannian Line Packing}
The set of all $n$-dimensional subspaces of $\mathbb{C}^N$ (or $\mathbb{R}^N$ for the real-valued case) is denoted as Grassmannian space $\mathcal{G}(N,n)$~\cite{ConwayHardinSloane1996}.
The problem of finding the best packing of $M$ $n$-dimensional subspaces in $\mathbb{C}^N$, with respect to some distance function, is commonly described as the Grassmannian subspace packing problem.
In literature, several distance functions are considered (e.g. the chordal or geodesic metric)~\cite{ConwayHardinSloane1996}.
For the one-dimensional case of $n=1$, which is also known as Grassmannian line packing, these metrics lead to the same optimal solution~\cite{dhilon_constructing_2008}.
This Grassmannian line packing corresponds to a Grassmannian frame (which motivated the name of these frames)~\cite{strohmer_grassmannian_2003}.

\subsection{Algorithms Aiming for Optimal Coherence}
The search for a vector set $\mathcal{A}$ with optimal coherence can be interpreted as optimization problem:
\begin{equation}
\label{eq:OptProb}
\argmin_{\mathcal{A}}\max \limits_{i\neq j} \frac{|\langle\boldsymbol{a}_i,\boldsymbol{a}_j\rangle|}{{{\|\boldsymbol{a}_i\|}}{{\|\boldsymbol{a}_j\|}}},\quad \boldsymbol{a}_i,\boldsymbol{a}_j\in\mathcal{A}
\end{equation}
Because of the importance for several research fields, a variety of algorithms has been developed in order to solve~\eqref{eq:OptProb} and to obtain optimal low-coherence vector sets.
Most algorithms aim for (nearly) MWBE sequences, however, if~\eqref{eq:WelchLimits} is not fulfilled, the other available bounds [e.g., \eqref{eq:orthoplex}, \eqref{eq:Levenshtein}, and \eqref{eq:BoundMukkavilli}] are targeted.

\subsubsection{Analytical approaches and direct solutions}
For several specific dimensions and numbers of vectors, there are analytical approaches to obtain MWBE sequences.
See~\cite{ConwayHardinSloane1996} for a summary of methods and solutions for the real-valued case.
A quite prominent example is based on conference matrices, allowing the construction of $N=M/2$ dimensional vector sets consisting of $M=p^\alpha+1$ vectors for the real-valued case and $M=2^{\alpha+1}$ vectors for the complex-valued case, where $p$ is an odd prime number and $\alpha\in\mathbb{N}$~\cite{LintSeidel1966,ConwayHardinSloane1996,HeathStrohmerPaulraj2003}.
There are also several other types of analytical approaches, for example sequences based on cosets of certain codes (e.g. expurgated sets of Gold sequences)~\cite{massey_welchs_1993,sarwate_meeting_1999} or the method of simplex signaling~\cite{proakis_digital_2008}. 
By extending an approach of~\cite{hochwald_systematic_2000}, where rows of an $N$-point Inverse Discrete Fourier Transform (IDFT) matrices are selected in order to build $\boldsymbol{A}$, cyclic difference sets are used in~\cite{xia_achieving_2005} to produce MWBE sequences.
This idea is further extended to different types of difference sets in~\cite{ding_generic_2007,ding_complex_2006}.
Another approach utilizes Steiner systems in order to build sparse ETFs~\cite{fickus_steiner_2012}.
It is shown in~\cite{jasper_kirkman_2014} that a large class of these Steiner ETFs can be transformed into so called Kirkman ETFs for which all entries are of constant modulus.

There are also analytical approaches for sequences which do not meet the Welch bound exactly, but are quite close to it (nearly MWBE).
For example, the previously mentioned MWBE approach based on difference sets can be extended to cyclotomic and almost difference sets~\cite{ding_complex_2006,ding_codebooks_2008,zhang_construction_2012,zhang_two_2012,hu_new_2014}.
By associated binary sequences, it can be even further generalized~\cite{yu_construction_2012,yu_new_2012_ISIT,yu_new_2012}.
Another approach for nearly MWBE sequences is based on the extended small Kasami codes or the non-linear Kerdock code~\cite{sarwate_meeting_1999}.

For the case of $M$ being too large to fulfill~\eqref{eq:WelchLimits}, the bound in~\eqref{eq:Levenshtein} is targeted by Mutually Unbiased Bases (MUBs) in~\cite{wootters_optimal_1989,ding_signal_2007}. MUBs consist of multiple $N$ dimensional bases for which all inner products across their elements are of the same magnitude $1/\sqrt{N}$~\cite{schwinger_unitary_1960,wootters_optimal_1989}.
Consequently, MUBs can also achieve the orthoplex bound~\eqref{eq:orthoplex}.
For $N$ being a prime power, the existence of MUBs with $N+1$ bases is shown in~\cite{wootters_optimal_1989}.
The construction provided therein is further generalized by~\cite{ding_signal_2007} and the corresponding matrices fulfill consequently the Levenshtein~\eqref{eq:Levenshtein} and the orthoplex bound~\eqref{eq:orthoplex} with equality.

\subsubsection{Numerical approaches}
If even existing, optimal solutions are only known for certain dimensions and numbers of vectors.
Thus, there is great interest for numerical approaches.
A variety of different algorithms have been published in the last decades and the whole topic is still active.
Examples are not limited to:
\begin{itemize}
\item Random search based DFT constructions leading to constellations with circulant structure~\cite{hochwald_systematic_2000}.
\item Considering a sphere vector quantizer obtained through a generalized Lloyd algorithm~\cite{xia_achieving_2005}.
\item Different smooth approximations of the $\max$ operator in~\eqref{eq:OptProb} with a free parameter which can be used in an iterative manner for optimization~\cite{agrawal_multiple-antenna_2001,gohary_noncoherent_2009,medra_flexible_2014}.
\item Application of an exponential map on space-time codes for coherent systems~\cite{kammoun_non-coherent_2007}.
\item Alternating projections enforcing alternately spectral and structural properties~\cite{dhilon_constructing_2008}.
\item A geometrically motivated expansion-compression algorithm~\cite{schober_geodesical_2009}.
\item Iterative decorrelation by a series of locally convex optimizations~\cite{rusu_design_2013}.
\item Combining shrinkage and matrix nearness with an optional averaging step~\cite{tsiligianni_construction_2014}.
\end{itemize}
In Section~\ref{sec:BCASC}, a new numerical approach is proposed, which aims to obtain vector sets with minimal coherence by finding BCASCs.

\subsection{Spherical Codes}
Any finite set of $M$ points placed on the surface of the $N$-dimensional unit sphere $\Omega_N$ centered at the origin of~$\mathbb{C}^N$ is called a spherical code and denoted by $C_s(N,M)$, where the suffix $(N,M)$ may later be skipped if it is of no further importance or clear from the context.
This definition is an extension to the non-complex variant from~\cite{ConwaySloane1999,EricsonZinoviev2001,SloaneWeb2}.
A point of $C_s(N,M) = \left\{ \boldsymbol{s}_m \right\}_{m=1}^M$ is determined by its position vector~$\boldsymbol{s}_m$ commonly interpreted as code word.
Therefore, a set of $M$ points can be equivalently described as $N \times M$ matrix. 
Best Spherical Codes (BSC), $C_{\mathrm{bs}}(N,M)$, are spherical codes which maximize the minimal Euclidean (or angular) distance $d_{ml}= \| \boldsymbol{s}_m - \boldsymbol{s}_l\|$ between any two points (or equivalently, minimize the maximal inner product of the corresponding vectors). 
All rotations of a BSC are usually regarded as the same, therefore, a BSC is characterized only by its distance distribution.

Considering also the complex antipodals within a spherical code, a CASC is denoted by $C_{cas}(N,M)$ and has the additional equivalence property:
\begin{equation}
\label{eq:equivalence}
\boldsymbol{s}_q\equiv\boldsymbol{s}_q\cdot e^{\mathrm{i}\phi} \in C_{cas}(N,M) \quad \forall \ \phi \in \mathbb{R},\  q \in \left\{1,\hdots, M\right\}
\end{equation}
Consequently, two equivalent vectors lie on the same line in~$\mathbb{C}^N$~\cite{LoveGrassmannians2003}. 
Every $C_{cas}(N,M)$ is obviously also a valid spherical code.
Naturally, a BCASC, denoted by $C_{bcas}$, also maximizes the minimal Euclidean distance between all its vectors, whereby the equivalence~\eqref{eq:equivalence} needs to be considered.

\subsection{Equivalence of Coherence and Distance Optimization}
\label{subsec:equivalence}
In order to use $C_{bcas}$ as low-coherence matrix, the equivalence of maximizing the minimal distance in a CASC and minimizing the coherence of a spherical code must be shown first.
This connection is commonly known, cf. \cite{LoveGrassmannians2003}, however, it is derived here explicitly for the sake of completeness and illustration.

Since $\left\vert \left< \boldsymbol{s}_p,\boldsymbol{s}_q\right>\right\vert=\left\vert\left< \boldsymbol{s}_p,\boldsymbol{s}_q\cdot e^{i\phi} \right>\right\vert \  \forall \ \boldsymbol{s}_p,\boldsymbol{s}_q \in C_s ,\ \phi \in \mathbb{R}$, we have
\begin{equation}
\min_{C_s} \max_{p\neq q} \left\vert \left< \boldsymbol{s}_p,\boldsymbol{s}_q \right>\right\vert
= \min_{C_{cas}} \max_{p\neq q} \left\vert \left<\boldsymbol{s}_p,\boldsymbol{s}_q\right>\right\vert.
\end{equation}
Therefore, it is sufficient to consider only $C_{cas}$ for the optimization. For these codes, the following holds as well:
\begin{align}
\Re\left( \left<\boldsymbol{s}_p,\boldsymbol{s}_q\right> \right)
\equiv\Im\left( \left<\boldsymbol{s}_p,\boldsymbol{s}_q\cdot e^{\mathrm{i}\frac{\pi}{2}}\right> \right)
\equiv\Im\left( \left<\boldsymbol{s}_p,\boldsymbol{s}_q\right> \right)
\end{align}
Consequently, it follows from the squared absolute of the inner product $\left\vert \left<\boldsymbol{s}_p,\boldsymbol{s}_q\right>\right\vert^2=\Re^2\left(\left< \boldsymbol{s}_p,\boldsymbol{s}_q\right> \right)+\Im^2\left( \left< \boldsymbol{s}_p,\boldsymbol{s}_q\right>\right)$:

\begin{equation}
\left\vert\left< \boldsymbol{s}_p,\boldsymbol{s}_q\right>\right\vert
\equiv\sqrt{2}\Re\left( \left<\boldsymbol{s}_p,\boldsymbol{s}_q\right> \right).
\end{equation}

Since $\left\| \boldsymbol{s}_p-\boldsymbol{s}_q\right\|^2=\left<\boldsymbol{s}_p,\boldsymbol{s}_p\right> + \left<\boldsymbol{s}_q,\boldsymbol{s}_q\right> - 2\Re\left(\left<\boldsymbol{s}_p,\boldsymbol{s}_q\right>\right)$,  the fact that the square function is monotonic for positive real values, and $\left<\boldsymbol{s}_p,\boldsymbol{s}_p\right> = \left<\boldsymbol{s}_q,\boldsymbol{s}_q\right>=1$, we have
\begin{equation}
C_{bcas}=\argmax_{C_{cas}}\min_{p\neq q}\left\| \boldsymbol{s}_p-\boldsymbol{s}_q\right\| 
=\argmin_{C_{s}} \max_{p\neq q} \left\vert \left< \boldsymbol{s}_p,\boldsymbol{s}_q \right>\right\vert.
\end{equation}
Thus, by finding a BCASC, a spherical code is obtained which results in a complex low-coherence matrix.

Note the importance of the CASC property~\eqref{eq:equivalence}:
E.g., for a CASC $\{\boldsymbol{s}_1=(1,0)^{\mathrm{T}},\boldsymbol{s}_2=(0,1)^{\mathrm{T}},\boldsymbol{s}_3=(\mathrm{i}\nicefrac{\sqrt{3}}{2},\nicefrac{1}{2})^{\mathrm{T}}\}$, the coherence is obtained for the pair $\boldsymbol{s}_1$ and $\boldsymbol{s}_3$ whilst the minimal distance is attained for the same pair with $\boldsymbol{s}_1\equiv\boldsymbol{s}_1\cdot \exp[{\mathrm{i}\nicefrac{\pi}{2}}]$.

\section{Best Complex Antipodal Spherical Codes}
\label{sec:BCASC}
In order to obtain codes which are close to $C_{bcas}$, the original method\cite{LazicSenkZamurovic1988} for obtaining codes close to $C_{bs}$ is adapted.
The new method can also be seen as a generalization of the Best Antipodal Spherical Codes (BASC) search approach~\cite{LazichZoerleinBossertSCC2013,ZoerleinLazichBossertSAMPTA2013}.
In this section, the principles of the original method\cite{LazicSenkZamurovic1988} are summarized and our approach for finding BCASCs is presented.

\subsection{Obtaining Best Spherical Codes}
In the following, the approach of \cite{LazicSenkZamurovic1988} for obtaining real-valued spherical codes which are typically very close to Best Spherical Codes is summarized and extended to complex sets.

The points of spherical codes can be considered as $M$ charged particles on the unit sphere acting in some field of repelling forces~\cite{Leech1957}. 
Starting from any initial position, such particles will move until the total potential energy of the system approaches some local minimum. 
In any one of these local minima, the particles will settle causing a stable or unstable equilibrium of mutual repelling forces.
In \cite{lazic_class_1980}, such a generalized potential function, $g(C_s(N,M))$, is introduced.
For a specific form of $g(C_s(N,M))$ given in \cite{lazic_construction_1986} by
\begin{equation}
	g(C_s(N,M)) = \sum\limits_{m=1}^M\sum\limits_{l<m} \| \boldsymbol{s}_m - \boldsymbol{s}_l\|^{-(\nu-2)},
	\label{eq:genpotfunc}
\end{equation}
where $\nu \in \mathbb{N}\  (\nu > 2)$, the global minimum of $g(C_s(N,M))$ is attained by a BSC if $\nu \rightarrow \infty$. 

Using the method of Lagrangian multipliers $\boldsymbol{\lambda}=\left\{ \lambda_m \right\}_{m=1}^M$ with $\lambda_m\in\mathbb{R}$, the Lagrange function $g(C_s(N,M),\boldsymbol{\lambda})$, corresponding to the potential function (\ref{eq:genpotfunc}) and the unit radius constraint of spherical codes, is given by
\begin{equation}
	g(C_s(N,M),\boldsymbol{\lambda}) = g(C_s(N,M)) + \sum\limits_{m=1}^{M} \lambda_m \left(\| \boldsymbol{s}_m\|^2 - 1\right).
	\label{eq:LagraneFunction}
\end{equation}
The necessary conditions for a global minimum of the potential function (\ref{eq:genpotfunc})
\begin{equation}
	 \frac{\partial g(C_s(N,M),\boldsymbol{\lambda})}{\partial s_{mn}} = 0 \quad  \mathrm{and} \quad  \frac{\partial g(C_s(N,M),\boldsymbol{\lambda})}{\partial \lambda_m} = 0,
\end{equation}
with  $m = 1,\hdots,M$ and $n = 1,\hdots,N$, can be expressed by the equilibrium (already derived in \cite{LazicSenkZamurovic1988}): 
\begin{equation}
	\left\{\boldsymbol{s}_m=\frac{\sum\limits_{l\neq m} \left[(\boldsymbol{s}_m - \boldsymbol{s}_l) / \| \boldsymbol{s}_m - \boldsymbol{s}_l \| ^\nu\right]}{ \left\| \sum\limits_{l\neq m} \left[ (\boldsymbol{s}_m - \boldsymbol{s}_l) / \| \boldsymbol{s}_m - \boldsymbol{s}_l \| ^{\nu}\right]\right\|} \right\}_{m=1}^M
\end{equation}
or, using hereafter the underlined denotation of unit vectors $\underline{\boldsymbol{u}}=\boldsymbol{u}/\|\boldsymbol{u}\|$, by 
\begin{equation}
	\left\{\underline{\boldsymbol{s}}_m=\underline{\sum\limits_{l\neq m} \frac{\underline{\boldsymbol{s}}_m - \underline{\boldsymbol{s}}_l }{ \| \underline{\boldsymbol{s}}_m - \underline{\boldsymbol{s}}_l \| ^\nu}} =\underline{\sum\limits_{l\neq m}\boldsymbol{\delta}_{ml}} \right\}_{m=1}^M.
	\label{eq:erdw}
\end{equation}

The right side of \eqref{eq:erdw} can be interpreted as collection of effective forces $\boldsymbol{f}_m$ acting on the code words of a spherical code.
By these forces, a mapping $\boldsymbol{P}$ can be introduced:
\begin{equation}
	\boldsymbol{P}[C_s(N,M)] = \left\{ \underline{\underline{\boldsymbol{s}}_m + \alpha \underline{\boldsymbol{f}}_m}\right\}_{m=1}^M,
	\label{eq:DampenedMapping}
\end{equation}
where $\underline{\boldsymbol{f}}_m$ is given by \eqref{eq:erdw} and $\alpha \in \mathbb{R}$.
For a small enough ``damping factor''~$\alpha$, the iterative process
\begin{equation}
	C_s(N,M)^{(k+1)} = \boldsymbol{P}(C_s(N,M)^{(k)}), \quad  k = 0, 1, \hdots
	\label{eq:iterative}
\end{equation}
converges to one fixed point of the function $\boldsymbol{P}$.

Already in~\cite{LazicSenkZamurovic1988}, it is numerically inferred that, generally, for $\nu$ large enough, all fixed points correspond to spherical codes whose minimal distances are close to the minimal distance of corresponding BSCs. 
Consequently by finding any fixed point using \eqref{eq:iterative} with $\nu$ large enough, the corresponding spherical code will be close to the best one with high probability.

The original description focused on real vector spaces. 
However, the approach is also valid for complex-valued case, since the global minimum of the generalized potential function $g(C_s(N,M))$ can be expressed by the equilibrium~\eqref{eq:erdw} also in the case of complex vector spaces.
For brevity, this is shown in the appendix.

For the case of an Euclidean space $\mathbb{R}^N$, the search for BSCs is adapted to the search of Grassmannian line packings by introducing the notion of BASC in~\cite{LazichZoerleinBossertSCC2013}.
In order to cover also the complex setting, BCASCs are proposed in the following.

\subsection{Obtaining Best Complex Antipodal Spherical Codes}
In order to consider the absolute value in the definition of the coherence \eqref{eq:coherence}, additional antipodal codewords resulting in BASC have been introduced in~\cite{LazichZoerleinBossertSCC2013} for the real-valued case.
Equivalent to this procedure, a factor of $\exp[\mathrm{i}\phi]$ [cf.~\eqref{eq:equivalence}] needs to be considered for the complex-valued case.
Similar to the approach in~\cite{LazichZoerleinBossertSCC2013}, this could be done by considering additional equivalent points.
However, the phase~$\phi$ in~\eqref{eq:equivalence} is continuous, and therefore, an infinite number of points would be needed.
In a first approximation, these infinite many points could be replaced by a finite number of $K$ distinct points generated by $\exp[\mathrm{i}2\pi k/K]\ \forall\  1 \leq k \leq K$ resulting in a new version of~\eqref{eq:erdw}:
\begin{equation}
\label{eq:SumApprox}
\left\{\underline{\boldsymbol{f}}_m
=\underline{\sum\limits_{k=1}^{K}\sum\limits_{l\neq m} \frac{\underline{\boldsymbol{s}}_m - \underline{\boldsymbol{s}}_l e^{\mathrm{i}2\pi\frac{k}{K}} }{ \left\| \underline{\boldsymbol{s}}_m - \underline{\boldsymbol{s}}_l e^{\mathrm{i}2\pi\frac{k}{K}} \right\| ^\nu}} \right\}_{m=1}^M
\end{equation}

As it is common for numerical approximations, the number of additional equivalent points must be large in order to have a valid approximation.
Consequently, if $K\rightarrow\infty$ is considered, the inserted sum in~\eqref{eq:SumApprox} becomes an integral:
\begin{equation}
\label{eq:Integral}
\left\{\underline{\boldsymbol{f}}_m
=\underline{\int\limits_{\kappa=0}^{2\pi}\sum\limits_{l\neq m} \frac{\underline{\boldsymbol{s}}_m - \underline{\boldsymbol{s}}_l e^{\mathrm{i}\kappa} }{ \left\| \underline{\boldsymbol{s}}_m - \underline{\boldsymbol{s}}_l e^{\mathrm{i}\kappa} \right\| ^\nu}d\kappa} \right\}_{m=1}^M
\end{equation}
Due to the norm in the denominator, this integral is hard to solve analytically (if this is even possible), therefore numerical integration must be used.

In the following, the described method will be denoted as BCASC search approach, since it arises from the quest for these special spherical codes.
However, it should be noted that this approach is not guaranteed to obtain BCASCs.

An algorithmic description of the BCASC search approach is given in Fig.\ref{fig:alg} altogether with exemplary parameters.
\begin{figure}
\algrenewcommand\algorithmicindent{1.3em}%
\begin{algorithmic}[1]
\State \textbf{procedure} \textsc{BCASC Search}($N$,$M$)
\State $\alpha_{\textrm{init}}\gets 0.9$, $\epsilon\gets 10^{-10}$ \Comment{numerical parameters}
\State {$\nu\gets2$, $\nu_{\max}\gets2^{10}$}
\State {$\tau_{\max}\gets10^5$, $\alpha\gets\alpha_{\textrm{init}}$}

\State {$C_s\gets$ random seed} \Comment{random spherical code}

\While{$ \nu < \nu_{\max}$}

    \State $\textrm{FixedPoint}\gets\mathbf{false}$ \Comment{initialize indicator}
    \State {$\tau\gets0$} \Comment{initialize iteration counter}
    
    \While{$\tau < \tau_{\max}$ \textbf{AND} $\textrm{FixedPoint}=\mathbf{false}$}

         \For{$m=1$ to $M$} \Comment{for each vector}
             \State  $\boldsymbol{f}_m
             \gets\int\limits_{\kappa=0}^{2\pi}\sum\limits_{l\neq m} \frac{\underline{\boldsymbol{s}}_m - \underline{\boldsymbol{s}}_l e^{\mathrm{i}\kappa} }{ \left\| \underline{\boldsymbol{s}}_m - \underline{\boldsymbol{s}}_l e^{\mathrm{i}\kappa} \right\| ^\nu}d\kappa$ \Comment{calc. forces}
		\EndFor
        \State  $\left.\left\{\boldsymbol{s}_m\right\}_{m=1}^M\right. \gets\left.\left\{\underline{\underline{\boldsymbol{s}}_m+\alpha\underline{\boldsymbol{f}}_m}\right\}_{m=1}^M\right. $ \Comment{apply forces}
		\If {all $\left\| \underline{\boldsymbol{f}}_m-\underline{\boldsymbol{s}}_m\right\| < \epsilon$} \Comment{check for fixed point}
			\State $\textrm{FixedPoint}\gets\mathbf{true}$ \Comment{stop loop and procede}
        \EndIf
		\State $\tau\gets \tau+1$
	\EndWhile
	\State $\nu\gets2\nu$ \Comment{adjust free parameter}
	\State $\alpha\gets\frac{\alpha_{\textrm{init}}}{\nu-1}$ \Comment{adjust damping factor}
    
\EndWhile
\State \Return {$A\gets\left\{ \underline{\boldsymbol{s}}_m\right\}_{m=1}^M$} \Comment{return obtained spherical code}
\State \textbf{end procedure}
\end{algorithmic}
\caption{\label{fig:alg}Iterative procedure for the search of BCASCs}
\end{figure}

\subsection{Best Complex Antipodal Codes in the Euclidean Space}
One should note that there is a slight technical difference in the way how an antipodal spherical code $C_{as}$  is defined in~\cite{LazichZoerleinBossertSCC2013} compared to the definition of $C_{cas}$ in this paper:
For the real-valued definition in~\cite{LazichZoerleinBossertSCC2013,ZoerleinLazichBossertSAMPTA2013}, $C_{as}$ contains both codewords, $\boldsymbol{s}_q$ and $-\boldsymbol{s}_q$, as individual vectors:
\begin{equation}
\boldsymbol{s}_m \in C_{\mathrm{as}}(N,M) \iff -\boldsymbol{s}_m \in C_{\mathrm{as}}(N,M)
\end{equation}
In contrast to this, the complex antipodal codewords are defined to be equivalent in the presented CASC-defini\-tion~\eqref{eq:equivalence}, and therefore, the cardinality of the vector set is equal to the number of non-collinear vectors.
However, the general spirit of utilizing antipodals remains the same for both approaches.
Consequently, the real-valued case can be treated as special case of the presented complex-valued case by a slightly different definition for antipodal spherical codes~$C_{as}$:
\begin{equation}
\boldsymbol{s}_q\equiv-\boldsymbol{s}_q \in C_{as}(N,M)  \quad \forall \  q \in \left\{1,\hdots, M\right\}
\end{equation}
In comparison to \eqref{eq:equivalence}, there is only one antipodal codeword which needs to be considered as equivalent for the real-valued case.
By this, the presented approach of defining antipodal codes over equivalence can be easily applied to the real-valued scenario as well.

\section{Discussion}
\label{sec:discussion}
\subsection{Near Optimal Solutions}
As stated before, it is not guaranteed that the global optimum is found by the presented approach. 
However, typically, the found fixed points are close to the global optimum.
We showed before in Section~\ref{subsec:equivalence} that the CASC with maximal minimum distance is also the vector set with minimal coherence.
However, the question remains to be answered whether a near optimal solution for a maximal minimum distance code also results in a near optimal solution for minimal coherence:
Due to $\left\| \boldsymbol{s}_p-\boldsymbol{s}_q\right\|^2=\left<\boldsymbol{s}_p,\boldsymbol{s}_p\right> + \left<\boldsymbol{s}_q,\boldsymbol{s}_q\right> - 2\Re\left(\left<\boldsymbol{s}_p,\boldsymbol{s}_q\right>\right)$, cf. Section~\ref{subsec:equivalence}, a discrepancy in the distance is squared while the inner product contributes only linear.
Therefore, this approach will also result in near optimal low-coherence solutions.

\subsection{Optimization Speedups}
\subsubsection{Accelerating on Straight Lines}
\label{sec:Accelerating}
As it is mentioned before, a sufficiently small ``damping factor''~$\alpha$ is necessary for convergence [see~\eqref{eq:DampenedMapping}].
However, a small value of $\alpha$ leads to slow convergence.
In order to cope with this situation, we adaptively determine the values of $\alpha$ on a per force basis:
If the direction of the force acting on some point $\underline{\boldsymbol{s}}_m$ has not changed from one iteration to another, the corresponding value of $\alpha$ is increased by a constant factor until a maximal value is reached. 
As an advantage, it is possible to start even with smaller values for $\alpha$ which lead naturally to preciser solutions.

\subsubsection{Numerical Integration}
\label{sec:Approx}
We used the QAG adaptive integration from the GSL~\cite{gough_gnu_2009} for the integral in~\eqref{eq:Integral}.
Since the integration needs to be performed for each step of the algorithm, this might be computationally expensive.
Approximating the integral simply by summing over $K$ points, as given in~\eqref{eq:SumApprox}, might be a fast alternative and is investigated in Section~\ref{sec:NumSimIntegralApprox}.
Since $M$ codewords interact with each other and the norm is calculated for each vector element with $K$ distinct points, the complexity of one iteration for a fixed $\nu$ scales asymptotically with $\mathcal{O}(N^2M^2K)$ in this case.

\section{Numerical Evaluation}
\label{sec:NumSim}
In this section, the presented method for coherence optimization is evaluated.
We used for our simulations the speedup mentioned in Section~\ref{sec:Accelerating}. The obtained matrices and the simulation files together with the corresponding parameters can be obtained from the first author.

\subsection{Coherence Optimization}
In order to evaluate the proposed optimization method, we selected the best coherence result obtained by optimization of ten random seeds, which are column-normalized matrices where the real and imaginary parts of the matrix elements are drawn from a standard normal distributed source.
For comparison, we also selected the best result out of ten runs for the implementation of the approach described by~\cite{medra_flexible_2014}, for which the source code can be found at \cite{medra_grassmannian_????}, as well as the lower bound given in~\eqref{eq:LowerBoundWhereWelchBoundFails} which includes the Welch bound.
In the figures, we added error bars indicating the worst, best, and average coherence found by the ten runs.

\begin{figure}[!t]
\centering
\begin{small}
\setlength\figurewidth{7.0cm}
\setlength\figureheight{7.0cm}
\begin{tikzpicture}

\begin{axis}[
width=\figurewidth,
height=\figureheight,
scale only axis,
xmin=4,
xmax=33,
xlabel={$M$},
xmajorgrids,
ymajorgrids,
ymin=0.3,
ymax=0.8,
ylabel={$\mu(\boldsymbol{A})$},
legend style={at={(0.97,0.03)},anchor=south east,draw=black,fill=white,legend cell align=left}
]

\addplot [
color=green!60!black,
only marks,
thick,
mark=asterisk,
mark options={solid}]
  table[row sep=crcr]{4	0.333333333343879\\
5	0.434340136109021\\
6	0.447213760051959\\
7	0.471404598376037\\
8	0.500011387606019\\
9	0.500017617713018\\
10	0.577350330973037\\
11	0.577350318950608\\
12	0.577350301740099\\
13	0.623921064710514\\
14	0.638154532723066\\
15	0.645071948751149\\
16	0.649094040716822\\
17	0.654752586274757\\
18	0.66332121460537\\
19	0.677722450947631\\
20	0.687149195142109\\
21	0.688163312386738\\
22	0.704347521927626\\
23	0.710251301686512\\
24	0.720226910545127\\
25	0.727199527194517\\
26	0.731333779971064\\
27	0.734820359979054\\
28	0.737356830329718\\
29	0.742323042516795\\
30	0.745519106927512\\
31	0.74596833093626\\
32	0.748870774921312\\
33	0.755280665912517\\
};
\addlegendentry{BCASC search};

\addplot [
color=orange!80!black,
only marks,
thick,
mark=o,
mark options={solid}
]
  table[row sep=crcr]{4	0.333470670551787\\
5	0.437089328281026\\
6	0.447447573671757\\
7	0.471450744270164\\
8	0.500341104601205\\
9	0.500189779806563\\
10	0.577395028478862\\
11	0.57742914281151\\
12	0.577431927331103\\
13	0.635913695828004\\
14	0.651821943695005\\
15	0.652037509956749\\
16	0.659005542873103\\
17	0.667904336515669\\
18	0.684183407655734\\
19	0.682524144485825\\
20	0.707050894551056\\
21	0.716943149411966\\
22	0.713422820177458\\
23	0.727269007098854\\
24	0.740122062198615\\
25	0.749019830333223\\
26	0.750289992606996\\
27	0.753309338727231\\
28	0.755399101134461\\
29	0.758657453732355\\
30	0.759208432011066\\
31	0.763988438998914\\
32	0.766182609914671\\
33	0.771218043124127\\
};
\addlegendentry{Approach by~\cite{medra_flexible_2014}};

\addplot [
color=blue,
thick,
densely dotted]
table[row sep=crcr]{4	0.333333333333333\\
5	0.408248290463863\\
6	0.447213595499958\\
7	0.471404520791032\\
8	0.487950036474267\\
9	0.5\\
10	0.577350269189626\\
11	0.577350269189626\\
12	0.577350269189626\\
13	0.591607978309962\\
14	0.603022689155527\\
15	0.612372435695794\\
16	0.620173672946042\\
17	0.626783170528009\\
18	0.632455532033676\\
19	0.637377439199098\\
20	0.641688947919748\\
21	0.645497224367903\\
22	0.64888568452305\\
23	0.651920240520265\\
24	0.654653670707977\\
25	0.657128740672771\\
26	0.659380473395787\\
27	0.661437827766148\\
28	0.66332495807108\\
29	0.665062171761176\\
30	0.666666666666667\\
31	0.668153104781061\\
32	0.669534063411986\\
33	0.670820393249937\\
};
\addlegendentry{Lower bound~\eqref{eq:LowerBoundWhereWelchBoundFails}};

\addplot [
color=blue,
thick,
solid]
table[row sep=crcr]{4	0.333333333333333\\
5	0.408248290463863\\
6	0.447213595499958\\
7	0.471404520791032\\
8	0.487950036474267\\
9	0.5\\
};

\addplot [color=green!80!black,only marks,mark=-,mark options={solid},forget plot,thick]
 plot [error bars/.cd, y dir = both, y explicit, error mark options={rotate=90, thick}, error bar style={thick}]
 table[row sep=crcr, y error plus index=2, y error minus index=3]{4	0.333333336107242	1.0687445206603e-08	2.76336320492732e-09\\
5	0.4343401592699	8.80889490795234e-08	2.31608788814341e-08\\
6	0.447213956058008	4.88664645192927e-07	1.96006049169029e-07\\
7	0.471404697917653	1.25238827197727e-07	9.95416164495389e-08\\
8	0.500012601949315	9.37853060434257e-07	1.21434329591885e-06\\
9	0.50001877574259	4.96878611699891e-07	1.15802957234568e-06\\
10	0.577350846082117	2.60945597352347e-07	5.15109079946008e-07\\
11	0.577350498950231	3.88770808990024e-07	1.79999623761695e-07\\
12	0.577350326374703	1.55245980515417e-08	2.46346045784662e-08\\
13	0.62438979192939	0.000538973710501267	0.000468727218876031\\
14	0.638154539553301	6.05126748709495e-09	6.83023515524184e-09\\
15	0.645094119318257	3.5849218621431e-06	2.21705671081818e-05\\
16	0.649094269933075	2.37749661513575e-07	2.29216253089248e-07\\
17	0.654752626321542	4.79381920825972e-08	4.00467847772035e-08\\
18	0.663321353440437	1.98432031917051e-08	1.38835067220455e-07\\
19	0.677730802358333	4.03639783674858e-06	8.35141070187895e-06\\
20	0.687149383171146	1.32352234283317e-07	1.88029037340165e-07\\
21	0.693447465715886	0.000587247825460269	0.00528415332914711\\
22	0.704347602061583	1.92440096302704e-07	8.01339573586191e-08\\
23	0.71025140140941	2.76740829363575e-07	9.97228984944698e-08\\
24	0.720227160346007	1.23039841315453e-07	2.49800880092188e-07\\
25	0.727300717907497	0.000451524097403677	0.000101190712980315\\
26	0.731468774825738	0.000666351849116786	0.000134994854674053\\
27	0.735562853296613	0.000181906818170097	0.000742493317558823\\
28	0.737757956159425	0.00160443416783573	0.000401125829707416\\
29	0.742454111901256	2.23688006163458e-05	0.000131069384460614\\
30	0.745928666511997	0.000614490661980049	0.00040955958448452\\
31	0.74933913707125	0.00227456421508065	0.00337080613499019\\
32	0.749548841369492	0.00520080124237698	0.000678066448180359\\
33	0.755372418241339	0.000820471455164062	9.17523288217481e-05\\
};
\addplot [color=orange!80!white,only marks,mark=-,mark options={solid},forget plot, thick]
 plot [error bars/.cd, y dir = both, y explicit, error mark options={rotate=90, thick}, error bar style={thick}]
 table[row sep=crcr, y error plus index=2, y error minus index=3]{4	0.333576249026572	9.09230229082514e-05	0.000105578474784984\\
5	0.437179852339127	0.000184770315340199	9.05240581010047e-05\\
6	0.447604967320321	0.000167016905267769	0.000157393648564041\\
7	0.495067751379758	0.00628054511780463	0.023617007109594\\
8	0.500586420163598	0.000240255723235272	0.000245315562393156\\
9	0.50034292992519	0.000191482254941255	0.000153150118627265\\
10	0.577432223004978	7.93057422263654e-05	3.71945261153916e-05\\
11	0.577460646900151	3.77162766787009e-05	3.15040886401663e-05\\
12	0.577446167736428	2.4211530758933e-05	1.4240405324939e-05\\
13	0.640227798253901	0.00345393186993648	0.00431410242589769\\
14	0.663805409138172	0.00272612350683066	0.0119834654431669\\
15	0.65385996352349	0.00566808536919505	0.00182245356674104\\
16	0.659599477390746	0.00084195624618788	0.000593934517643224\\
17	0.670150260189456	0.00223281253121255	0.00224592367378751\\
18	0.684981981975199	0.000500379530547823	0.000798574319464951\\
19	0.682630506334914	0.000402679717602106	0.00010636184908841\\
20	0.707412232230524	0.000534016308392227	0.000361337679468443\\
21	0.718802366721328	0.0024695617057281	0.00185921730936212\\
22	0.715014297290924	0.00237691748318802	0.00159147711346619\\
23	0.729509216369186	0.00237129439472006	0.00224020927033231\\
24	0.743255796916365	0.00274411491627213	0.00313373471775036\\
25	0.751330547521797	0.00360123961656544	0.0023107171885739\\
26	0.754950188562537	0.00729549367161642	0.0046601959555409\\
27	0.756580375376041	0.00559293081877288	0.00327103664880979\\
28	0.75987817291708	0.00261997391998836	0.00447907178261886\\
29	0.760382939987075	0.0013214278630298	0.00172548625472047\\
30	0.761501206614214	0.00588023658794268	0.00229277460314736\\
31	0.766602795973997	0.0069689891745871	0.00261435697508272\\
32	0.770840321679247	0.00452414319035743	0.0046577117645763\\
33	0.773081265505545	0.00457971963892612	0.00186322238141834\\
};
\end{axis}
\end{tikzpicture}
\end{small}
\caption{\label{fig:Dim3}Best obtained coherence out of ten runs for each number of vectors~$M$ in N=3 dimensions. The proposed BCASC search method is compared to the approach in~\cite{medra_flexible_2014}. The errorbars indicate the variance in the ten runs. The line for the lower bound is solid if~\eqref{eq:WelchLimits} is fulfilled.}
\end{figure}
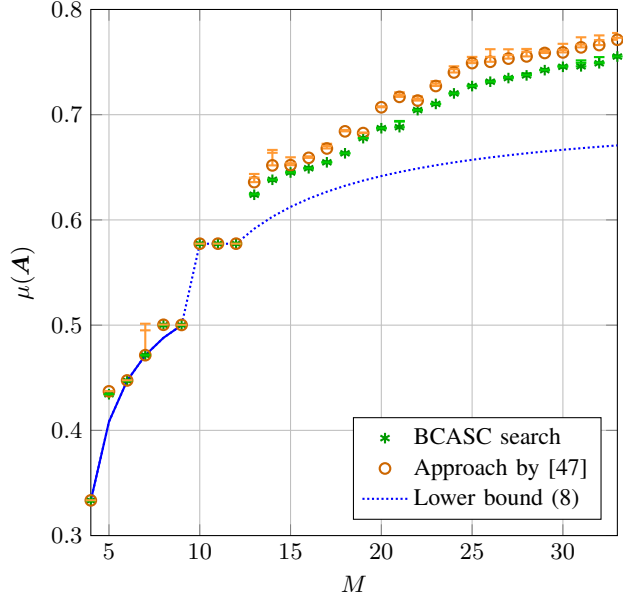

For $N=3$, a comparison of the obtained coherence values is illustrated in Fig.~\ref{fig:Dim3}.
As long as the Welch bound can be met according to~\eqref{eq:WelchLimits} ($M\leq 9=N^2$), the coherence of our vector sets actually meets the Welch bound with equality or is very close to it.
For $M\leq 12$, our BCASC search method and the approach by~\cite{medra_flexible_2014} yield almost equivalent results.
Observe the quasi-constant coherence level obtained with both methods for $10\leq M \leq 12$.
This is also observed in other dimensions, where it will be discussed later, and corresponds to the orthoplex bound~\eqref{eq:orthoplex}.
For sets with more vectors, our method obtains smaller coherence levels.
As it can be seen from the error bars in the simulation results, our proposed method produces stable optimization results.
Therefore, most of the found (local) optima yield similar coherence values.
Starting from $M\geq13$, the orthoplex bound is replaced by~\eqref{eq:Levenshtein} as active lower bound in~\eqref{eq:LowerBoundWhereWelchBoundFails}.
For even larger vector sets, the bound of~\eqref{eq:BoundMukkavilli} is dominating for $M\geq39$.
As it is stated in~\cite{xia_achieving_2005}, the lower bound~\eqref{eq:BoundMukkavilli} gets tighter for larger~$M$.

\begin{figure*}[!t]
\centering
\subfloat[$N=4$]
{
	\begin{small}
	\setlength\figurewidth{6.0cm}
	\setlength\figureheight{6.0cm}
	\begin{tikzpicture}
	
	\begin{axis}[
	width=\figurewidth,
	height=\figureheight,
	scale only axis,
	xmin=5,
	xmax=51,
	xlabel={$M$},
	xmajorgrids,
	ymajorgrids,
	ymin=0.25,
	ymax=0.75,
	ylabel={$\mu(\boldsymbol{A})$},
	legend style={at={(0.97,0.03)},anchor=south east,draw=black,fill=white,legend cell align=left}
	]

	\addplot [
	color=green!60!black,
	only marks,
	thick,
	mark=asterisk,
	mark options={solid}]
	  table[row sep=crcr]{5	0.250000007569972\\
	  6	0.327749857585534\\
	  7	0.353553627872307\\
	  8	0.377969126003591\\
	  9	0.402163189392958\\
	  10	0.411802604162623\\
	  11	0.425851201962673\\
	  12	0.4280204249813\\
	  13	0.433013345199943\\
	  14	0.447728585021628\\
	  15	0.447438302772921\\
	  16	0.447213711549423\\
	  17	0.50000018537678\\
	  18	0.500000500536609\\
	  19	0.500000268585246\\
	  20	0.500000243981297\\
	  21	0.538466020292316\\
	  22	0.546325453946564\\
	  23	0.553381186252494\\
	  24	0.560311841801383\\
	  25	0.567340827728335\\
	  26	0.572814629229758\\
	  27	0.577064627241962\\
	  28	0.577935201538345\\
	  29	0.577815419033884\\
	  30	0.577821295566842\\
	  31	0.577901023176634\\
	  32	0.577909726789335\\
	  33	0.577986822077266\\
	  34	0.577721473785273\\
	  35	0.578065058352985\\
	  36	0.577809917122509\\
	  37	0.577668640860349\\
	  38	0.577542196134304\\
	  39	0.577434373487968\\
	  40	0.577350325958916\\
	  41	0.622707978450431\\
	  42	0.625895060972588\\
	  43	0.630561733899173\\
	  44	0.632038731286787\\
	  45	0.63427220340435\\
	  46	0.636760669558764\\
	  47	0.641151603039691\\
	  48	0.644963811323624\\
	  49	0.648250366046711\\
	  50	0.650771158764576\\
	  51	0.65415619528672\\
	};
	\addlegendentry{BCASC search};
	
	\addplot [
	color=orange!80!black,
	only marks,
	thick,
	mark=o,
	mark options={solid}
	]
	  table[row sep=crcr]{5	0.250181070132959\\
	  6	0.327441266214055\\
	  7	0.354028393104991\\
	  8	0.378679143435015\\
	  9	0.402106827126932\\
	  10	0.411324887147586\\
	  11	0.428136943397243\\
	  12	0.42786705816167\\
	  13	0.448477838715774\\
	  14	0.447867837182685\\
	  15	0.447462428400211\\
	  16	0.447305039967579\\
	  17	0.500097818259098\\
	  18	0.500162250116184\\
	  19	0.500118590546587\\
	  20	0.500126628338602\\
	  21	0.559910380221487\\
	  22	0.563895839237374\\
	  23	0.564734065343967\\
	  24	0.580828973120422\\
	  25	0.59014172553701\\
	  26	0.591545179080529\\
	  27	0.594764688825176\\
	  28	0.597841083522828\\
	  29	0.608225380883528\\
	  30	0.605334456200913\\
	  31	0.605295923749988\\
	  32	0.603568981057645\\
	  33	0.60079062050155\\
	  34	0.595415960644374\\
	  35	0.604050238047468\\
	  36	0.59800665536734\\
	  37	0.594066757279291\\
	  38	0.58783595683328\\
	  39	0.583678892132629\\
	  40	0.577509249682886\\
	  41	0.678677463579892\\
	  42	0.685080481413332\\
	  43	0.683214854601504\\
	  44	0.65164112305324\\
	  45	0.652054502822124\\
	  46	0.66185190695014\\
	  47	0.667965507027669\\
	  48	0.669045791387816\\
	  49	0.670954256255872\\
	  50	0.676714319555256\\
	  51	0.681563637269909\\
	};
	\addlegendentry{Approach by~\cite{medra_flexible_2014}};
	
	\addplot [
	color=blue,
	thick,
	densely dotted]
	table[row sep=crcr]{5	0.25\\
	6	0.316227766016838\\
	7	0.353553390593274\\
	8	0.377964473009227\\
	9	0.395284707521047\\
	10	0.408248290463863\\
	11	0.418330013267038\\
	12	0.426401432711221\\
	13	0.433012701892219\\
	14	0.438529009653515\\
	15	0.443202630213959\\
	16	0.447213595499958\\
	17	0.5\\
	18	0.5\\
	19	0.5\\
	20	0.5\\
	21	0.508747019069168\\
	22	0.516397779494322\\
	23	0.523148363780597\\
	24	0.529150262212918\\
	25	0.534522483824849\\
	26	0.539359889970594\\
	27	0.543739067856121\\
	28	0.547722557505166\\
	29	0.551361950083609\\
	30	0.554700196225229\\
	31	0.557773351022717\\
	32	0.560611910581388\\
	33	0.563241847975046\\
	34	0.565685424949238\\
	35	0.567961834247065\\
	36	0.570087712549569\\
	37	0.572077553547355\\
	38	0.573944043183551\\
	39	0.575698333703139\\
	40	0.577350269189626\\
	41	0.578908572345526\\
	42	0.580381000088009\\
	43	0.58177447388274\\
	44	0.58309518948453\\
	45	0.584348709790778\\
	46	0.58554004376912\\
	47	0.58667371384069\\
	48	0.587753813645259\\
	49	0.58878405775519\\
	50	0.589767824619589\\
	51	0.590708193791796\\
	};
	\addlegendentry{Lower bound~\eqref{eq:LowerBoundWhereWelchBoundFails}};
	
	\addplot [
	color=blue,
	thick,
	solid]
	table[row sep=crcr]{5	0.25\\
	6	0.316227766016838\\
	7	0.353553390593274\\
	8	0.377964473009227\\
	9	0.395284707521047\\
	10	0.408248290463863\\
	11	0.418330013267038\\
	12	0.426401432711221\\
	13	0.433012701892219\\
	14	0.438529009653515\\
	15	0.443202630213959\\
	16	0.447213595499958\\
	};
	
	\addplot [color=green!80!black,only marks,mark=-,mark options={solid},forget plot,thick]
	 plot [error bars/.cd, y dir = both, y explicit, error mark options={rotate=90, thick}, error bar style={thick}]
	 table[row sep=crcr, y error plus index=2, y error minus index=3]{5	0.250000055657903	3.00981967127356e-08	4.80879311370863e-08\\
	 6	0.327749859120402	8.94184004618381e-10	1.53486739984388e-09\\
	 7	0.353553742735132	8.51897833387127e-08	1.14862825306705e-07\\
	 8	0.377971061629969	1.86335066454379e-06	1.93562637801348e-06\\
	 9	0.402163314365244	3.43414960357613e-07	1.2497228607522e-07\\
	 10	0.411802684005261	8.50136238073063e-08	7.98426388337603e-08\\
	 11	0.426211637984985	0.000360249395654211	0.000360436022312671\\
	 12	0.428020525832917	3.5383159707747e-07	1.0085161661344e-07\\
	 13	0.433013568495588	7.46841048282487e-07	2.23295644918853e-07\\
	 14	0.452414616629406	0.0020091440822117	0.00468603160777786\\
	 15	0.447438349389794	4.17100806227033e-07	4.66168729174044e-08\\
	 16	0.447214023526532	8.19910034643989e-07	3.11977109201944e-07\\
	 17	0.500415459505034	0.000148320531415669	0.000415274128253817\\
	 18	0.500467397745902	0.00419935954764361	0.000466897209292494\\
	 19	0.500000484678482	1.93242186918852e-07	2.16093235638759e-07\\
	 20	0.521668025781997	0.00686654366928574	0.0216677818006999\\
	 21	0.538766182356728	0.00116544483050762	0.000300162064411968\\
	 22	0.547395943641714	0.00135421459766771	0.00107048969514989\\
	 23	0.554864461383246	0.00292190070839149	0.00148327513075197\\
	 24	0.560916194847503	6.7421064266826e-05	0.000604353046119832\\
	 25	0.569065492379013	0.000863546200621013	0.00172466465067855\\
	 26	0.573459943812454	0.000816050326374218	0.000645314582696455\\
	 27	0.577826409607156	0.00132600034747221	0.000761782365194197\\
	 28	0.583109661181275	0.000838035873993004	0.0051744596429304\\
	 29	0.581326765709613	0.00798839420232045	0.00351134667572905\\
	 30	0.577837079694564	5.17662628423121e-05	1.57841277214077e-05\\
	 31	0.577959049045372	0.000146680240133978	5.8025868737932e-05\\
	 32	0.58033565073459	0.0217191062547742	0.00242592394525565\\
	 33	0.577995448082826	7.70238575179549e-05	8.62600556006399e-06\\
	 34	0.577750566365768	0.000261137273613987	2.90925804946607e-05\\
	 35	0.578065188208644	4.36650532842364e-07	1.2985565922552e-07\\
	 36	0.577844728284421	3.47718523631135e-05	3.48111619119518e-05\\
	 37	0.577692247461513	3.7463683947303e-06	2.36066011645164e-05\\
	 38	0.577542694433702	1.70165208068784e-06	4.98299397411195e-07\\
	 39	0.577435201401363	1.31529459068691e-06	8.27913395173852e-07\\
	 40	0.5773503514786	6.71265959484302e-08	2.55196834775262e-08\\
	 41	0.623314971532214	0.00141591157879351	0.000606993081782803\\
	 42	0.625959525106219	0.000242279038297544	6.44641336305929e-05\\
	 43	0.630841934396609	0.000630709249105355	0.000280200497435934\\
	 44	0.633350267626941	0.0024665810138581	0.00131153634015457\\
	 45	0.63596474134358	0.00222746801966067	0.0016925379392293\\
	 46	0.638707661173554	0.00111782738606792	0.00194699161478973\\
	 47	0.643056080820932	0.00114125277903432	0.00190447778124014\\
	 48	0.645855325774864	0.00110042012688605	0.000891514451239361\\
	 49	0.649359998946647	0.000882492275753322	0.00110963289993593\\
	 50	0.65198211348404	0.00166604252006142	0.00121095471946464\\
	 51	0.655029388776023	0.000738883326439677	0.000873193489302149\\
	};
	\addplot [color=orange!80!white,only marks,mark=-,mark options={solid},forget plot, thick]
	 plot [error bars/.cd, y dir = both, y explicit, error mark options={rotate=90, thick}, error bar style={thick}]
	 table[row sep=crcr, y error plus index=2, y error minus index=3]{5	0.250256500095089	8.59361244953671e-05	7.5429962130058e-05\\
	 6	0.327562288957829	0.000170273456029768	0.00012102274377368\\
	 7	0.35438335561245	0.000735748244774515	0.0003549625074587\\
	 8	0.37876886821123	8.78112920263963e-05	8.97247762146614e-05\\
	 9	0.404466749471101	0.00373487227373309	0.00235992234416882\\
	 10	0.412108617236772	0.00624018674012938	0.000783730089185863\\
	 11	0.429754879659156	0.00481317511136742	0.00161793626191276\\
	 12	0.443348583206181	0.00732812713814512	0.015481525044511\\
	 13	0.448692212974577	0.000450111947389809	0.000214374258803163\\
	 14	0.447986701821179	0.000145820803980345	0.000118864638493921\\
	 15	0.447505598900723	4.08700227611525e-05	4.31705005119465e-05\\
	 16	0.447324070157083	1.23313957008375e-05	1.90301895046097e-05\\
	 17	0.506436370793944	0.00743455884893018	0.00633855253484583\\
	 18	0.512536194605915	0.0106039016651999	0.0123739444897306\\
	 19	0.500155124287144	3.1110114525168e-05	3.65337405567878e-05\\
	 20	0.538583760979757	0.00632223869555182	0.038457132641155\\
	 21	0.564091861286217	0.0153893138456752	0.00418148106473004\\
	 22	0.569131893493934	0.00581626254272782	0.00523605425656026\\
	 23	0.574880790541963	0.00355233286332379	0.0101467251979961\\
	 24	0.582712197904718	0.0041083108859854	0.00188322478429659\\
	 25	0.59353365549867	0.00555794902968498	0.00339192996166005\\
	 26	0.593482940265656	0.00368982924221239	0.00193776118512745\\
	 27	0.598943533630651	0.00171018201730322	0.00417884480547526\\
	 28	0.605266055914309	0.00574848455921462	0.00742497239148165\\
	 29	0.610921929066696	0.0029783980142043	0.00269654818316811\\
	 30	0.608315051854438	0.00446478657784188	0.00298059565352526\\
	 31	0.610141991727189	0.0128024656649894	0.00484606797720122\\
	 32	0.607580475908014	0.00645406149149641	0.00401149485036889\\
	 33	0.607601678982327	0.011865335143275	0.00681105848077712\\
	 34	0.601589971232409	0.0193342450322245	0.00617401058803524\\
	 35	0.606268617905021	0.00586898805438651	0.00221837985755291\\
	 36	0.607905159471295	0.00827723751105558	0.00989850410395521\\
	 37	0.607041864065324	0.0127605938114399	0.0129751067860331\\
	 38	0.591646585671498	0.00362061118379087	0.00381062883821792\\
	 39	0.583709332101183	2.12187059555058e-05	3.04399685547851e-05\\
	 40	0.577540801027125	2.34445734104138e-05	3.15513442386006e-05\\
	 41	0.684515061369422	0.00851398013323834	0.00583759778952953\\
	 42	0.693282439971474	0.00882511824241106	0.0082019585581421\\
	 43	0.685612467527145	0.0163285212379751	0.002397612925641\\
	 44	0.662902989570334	0.0261755307708376	0.0112618665170942\\
	 45	0.660228155982808	0.0238621441899256	0.00817365316068364\\
	 46	0.674520327550319	0.00937102522597966	0.0126684206001799\\
	 47	0.677144881466732	0.00777890609683329	0.00917937443906292\\
	 48	0.676211639408202	0.00731260841515524	0.00716584802038589\\
	 49	0.678647015497252	0.0043466109376612	0.00769275924138013\\
	 50	0.680496512741255	0.00270863492001971	0.0037821931859997\\
	 51	0.683193966120436	0.00253472312593206	0.0016303288505265\\
	};
	\end{axis}
	\end{tikzpicture}
	\end{small}
	\label{fig:Dim4}
}
\hfil
\subfloat[$N=5$]
{
	\begin{small}
	\setlength\figurewidth{6.0cm}
	\setlength\figureheight{6.0cm}
	\begin{tikzpicture}
	
	\begin{axis}[
	width=\figurewidth,
	height=\figureheight,
	scale only axis,
	xmin=6,
	xmax=57,
	xlabel={$M$},
	xmajorgrids,
	ymajorgrids,
	ymin=0.2,
	ymax=0.65,
	ylabel={$\mu(\boldsymbol{A})$},
	legend style={at={(0.97,0.03)},anchor=south east,draw=black,fill=white,legend cell align=left}
	]

	\addplot [
	color=green!60!black,
	only marks,
	thick,
	mark=asterisk,
	mark options={solid}]
	  table[row sep=crcr]{6	0.200000037116076\\
	  7	0.266967245614883\\
	  8	0.295477961534835\\
	  9	0.320668676128978\\
	  10	0.333334116249026\\
	  11	0.346510190729837\\
	  12	0.357919728406922\\
	  13	0.368628965367967\\
	  14	0.376291293159787\\
	  15	0.380437615005229\\
	  16	0.388945608628708\\
	  17	0.39441983515894\\
	  18	0.401985408107897\\
	  19	0.39817322204722\\
	  20	0.397777684225906\\
	  21	0.400000718470308\\
	  22	0.408794227009744\\
	  23	0.408700847954035\\
	  24	0.408431451753415\\
	  25	0.408248298967925\\
	  26	0.447561886470188\\
	  27	0.454650187895806\\
	  28	0.44768843035545\\
	  29	0.447213967754279\\
	  30	0.447213707381682\\
	  31	0.483464375166366\\
	  32	0.488485969790022\\
	  33	0.494284555398855\\
	  34	0.498739691699115\\
	  35	0.502692960804016\\
	  36	0.509037395598889\\
	  37	0.513260713938618\\
	  38	0.517775564180972\\
	  39	0.500806347794818\\
	  40	0.524955954302791\\
	  41	0.50047496779369\\
	  42	0.528460236529818\\
	  43	0.500331414635334\\
	  44	0.521164623294052\\
	  45	0.521452866322031\\
	  46	0.536582429518175\\
	  47	0.539008922305572\\
	  48	0.543232747808078\\
	  49	0.546500660217441\\
	  50	0.546836234374366\\
	  51	0.553866966084999\\
	  52	0.556288539481679\\
	  53	0.557888784284458\\
	  54	0.559900290504866\\
	  55	0.554911732988506\\
	  56	0.558591861693518\\
	  57	0.559394368713456\\
	};
	\addlegendentry{BCASC search};
	
	\addplot [
	color=orange!80!black,
	only marks,
	thick,
	mark=o,
	mark options={solid}
	]
	  table[row sep=crcr]{6	0.20017652212937\\
	  7	0.266476162182601\\
	  8	0.295412142558909\\
	  9	0.320346829029762\\
	  10	0.334084027090715\\
	  11	0.347288518474197\\
	  12	0.358151795788938\\
	  13	0.369094823000506\\
	  14	0.378782017363333\\
	  15	0.385755042252513\\
	  16	0.393235981304498\\
	  17	0.398995064079408\\
	  18	0.4099067685086\\
	  19	0.400658162893536\\
	  20	0.397765775767504\\
	  21	0.409027221401592\\
	  22	0.408722184830236\\
	  23	0.408544771924701\\
	  24	0.408406167108212\\
	  25	0.408370854096087\\
	  26	0.447337662002291\\
	  27	0.474213602829194\\
	  28	0.470260368389659\\
	  29	0.447373249815547\\
	  30	0.447343676054475\\
	  31	0.505176160718257\\
	  32	0.511264935188586\\
	  33	0.515050554427791\\
	  34	0.523238974615056\\
	  35	0.527783726806706\\
	  36	0.531325963368285\\
	  37	0.535274696099636\\
	  38	0.534998998891181\\
	  39	0.544857314139546\\
	  40	0.53199416705093\\
	  41	0.546836188988277\\
	  42	0.5563166575966\\
	  43	0.558301665383133\\
	  44	0.539763290664118\\
	  45	0.566166913021962\\
	  46	0.567382641298067\\
	  47	0.570866427229419\\
	  48	0.574448359575566\\
	  49	0.576016493001734\\
	  50	0.579132562282991\\
	  51	0.581059921338003\\
	  52	0.58338608170041\\
	  53	0.587449375963846\\
	  54	0.587337049726829\\
	  55	0.588542167026065\\
	  56	0.590501284238095\\
	  57	0.593192569614729\\
	};
	\addlegendentry{Approach by~\cite{medra_flexible_2014}};
	
	\addplot [
	color=blue,
	thick,
	densely dotted]
	table[row sep=crcr]{6	0.2\\
	7	0.258198889747161\\
	8	0.29277002188456\\
	9	0.316227766016838\\
	10	0.333333333333333\\
	11	0.346410161513775\\
	12	0.356753034006338\\
	13	0.365148371670111\\
	14	0.372104203767625\\
	15	0.377964473009227\\
	16	0.382970843102535\\
	17	0.387298334620742\\
	18	0.391076944437521\\
	19	0.394405318873308\\
	20	0.397359707119513\\
	21	0.4\\
	22	0.402373908081478\\
	23	0.404519917477945\\
	24	0.40646942234853\\
	25	0.408248290463863\\
	26	0.447213595499958\\
	27	0.447213595499958\\
	28	0.447213595499958\\
	29	0.447213595499958\\
	30	0.447213595499958\\
	31	0.452910813657838\\
	32	0.458122847290851\\
	33	0.462910049886276\\
	34	0.467323019546118\\
	35	0.471404520791032\\
	36	0.475190963311491\\
	37	0.478713553878169\\
	38	0.481999203654147\\
	39	0.485071250072666\\
	40	0.487950036474267\\
	41	0.490653381462658\\
	42	0.493196961916072\\
	43	0.495594627783352\\
	44	0.497858662537118\\
	45	0.5\\
	46	0.502028405894729\\
	47	0.50395263067897\\
	48	0.505780538858873\\
	49	0.507519218922552\\
	50	0.509175077217316\\
	51	0.510753918455249\\
	52	0.512261015028209\\
	53	0.513701166914081\\
	54	0.515078753637713\\
	55	0.516397779494322\\
	56	0.517661913037023\\
	57	0.518874521662771\\
	};
	\addlegendentry{Lower bound~\eqref{eq:LowerBoundWhereWelchBoundFails}};
	
	\addplot [
	color=blue,
	thick,
	solid]
	table[row sep=crcr]{6	0.2\\
	7	0.258198889747161\\
	8	0.29277002188456\\
	9	0.316227766016838\\
	10	0.333333333333333\\
	11	0.346410161513775\\
	12	0.356753034006338\\
	13	0.365148371670111\\
	14	0.372104203767625\\
	15	0.377964473009227\\
	16	0.382970843102535\\
	17	0.387298334620742\\
	18	0.391076944437521\\
	19	0.394405318873308\\
	20	0.397359707119513\\
	21	0.4\\
	22	0.402373908081478\\
	23	0.404519917477945\\
	24	0.40646942234853\\
	25	0.408248290463863\\
	};
	
	\addplot [color=green!80!black,only marks,mark=-,mark options={solid},forget plot,thick]
	 plot [error bars/.cd, y dir = both, y explicit, error mark options={rotate=90, thick}, error bar style={thick}]
	 table[row sep=crcr, y error plus index=2, y error minus index=3]{6	0.20000007876045	3.25310764714981e-08	4.16443740880457e-08\\
	 7	0.266967539189149	1.387975251399e-07	2.93574266729824e-07\\
	 8	0.295478106478558	2.68267702330238e-07	1.44943722857338e-07\\
	 9	0.320668746554345	6.34819287137489e-08	7.0425366427429e-08\\
	 10	0.333366300892356	3.89059901507172e-05	3.21846433300466e-05\\
	 11	0.346510400507028	1.66036405291337e-07	2.09777191229055e-07\\
	 12	0.358199097665619	7.82629174008176e-05	0.000279369258696816\\
	 13	0.36885324245823	0.000443665125020476	0.000224277090262903\\
	 14	0.376993454687187	0.000492021306250201	0.000702161527400047\\
	 15	0.382812152780529	0.00133085397928373	0.00237453777529972\\
	 16	0.389960734891833	0.00329452575266376	0.00101512626312544\\
	 17	0.394860183414694	0.00176366584504939	0.000440348255753342\\
	 18	0.402406336692101	0.000311265617095435	0.000420928584204061\\
	 19	0.399301136808779	0.01012474797423	0.00112791476155943\\
	 20	0.39889331648007	0.0100337169221753	0.00111563225416372\\
	 21	0.411095433087911	0.00392701161925524	0.0110947146176024\\
	 22	0.408826461665286	8.84630050976831e-05	3.22346555419739e-05\\
	 23	0.408704538959246	6.0745814703389e-06	3.69100521024901e-06\\
	 24	0.408431518670418	2.89394115893238e-08	6.6917003105349e-08\\
	 25	0.412021481208711	0.0150920555171354	0.00377318224078643\\
	 26	0.447562278910829	8.78318593466876e-07	3.9244064092836e-07\\
	 27	0.456016926540741	0.00126831738246996	0.00136673864493542\\
	 28	0.457303270136804	0.00399860646777273	0.00961483978135369\\
	 29	0.447345881847205	5.88365955005998e-05	0.000131914092925844\\
	 30	0.447214017997929	2.05523955743026e-07	3.10616247001239e-07\\
	 31	0.484591876301677	0.00105399998416189	0.00112750113531046\\
	 32	0.489576332794987	0.00125798892087281	0.00109036300496557\\
	 33	0.495274144033886	0.00114013279722608	0.000989588635031513\\
	 34	0.500345872837161	0.00139068815750476	0.00160618113804545\\
	 35	0.504867845201223	0.00115078416087055	0.0021748843972067\\
	 36	0.509736168752138	0.000651733740070215	0.00069877315324951\\
	 37	0.513972741408645	0.000906276992333699	0.000712027470027188\\
	 38	0.518454518619694	0.000477026686313775	0.000678954438721902\\
	 39	0.520118279707443	0.00318565151050931	0.0193119319126248\\
	 40	0.525814814644444	0.000363998291172507	0.0008588603416535\\
	 41	0.525023059623615	0.00509239401932493	0.0245480918299249\\
	 42	0.532559123906181	0.00200970082923979	0.00409888737636244\\
	 43	0.532370016791855	0.00446991714711364	0.0320386021565209\\
	 44	0.533343988969519	0.00614538881700788	0.0121793656754667\\
	 45	0.53221716743101	0.00971678811849619	0.010764301108979\\
	 46	0.54340235342906	0.00186512885398382	0.00681992391088471\\
	 47	0.544348374828502	0.00353788128716037	0.00533945252293011\\
	 48	0.54796497412343	0.00123314948184938	0.00473222631535164\\
	 49	0.550543962411384	0.00190760983758798	0.00404330219394278\\
	 50	0.552736102406174	0.00180170199600516	0.00589986803180753\\
	 51	0.554954872971878	0.00176283130046762	0.00108790688687943\\
	 52	0.557444877732921	0.00069964938859568	0.00115633825124117\\
	 53	0.558936533956825	0.00166719404650517	0.00104774967236632\\
	 54	0.561041540169776	0.000790188648834289	0.00114124966491036\\
	 55	0.561948934413357	0.00157265236197901	0.00703720142485165\\
	 56	0.563328396129407	0.00233586561070021	0.00473653443588862\\
	 57	0.564208834115331	0.00128254416848272	0.00481446540187513\\
	};
	\addplot [color=orange!80!white,only marks,mark=-,mark options={solid},forget plot, thick]
	 plot [error bars/.cd, y dir = both, y explicit, error mark options={rotate=90, thick}, error bar style={thick}]
	 table[row sep=crcr, y error plus index=2, y error minus index=3]{6	0.200239734090462	4.28388401354507e-05	6.32119610917692e-05\\
	 7	0.266630591942082	0.000115447452995354	0.000154429759481323\\
	 8	0.295892315493262	0.000241063932447771	0.000480172934352985\\
	 9	0.320956353487385	0.00171129331000675	0.000609524457623523\\
	 10	0.334597387277524	0.00116251891747765	0.000513360186808365\\
	 11	0.349173985853921	0.0021171325809497	0.00188546737972339\\
	 12	0.360164444230599	0.00357290975480301	0.0020126484416606\\
	 13	0.371507376004138	0.0047803137905888	0.00241255300363147\\
	 14	0.380707637169888	0.00329401339064384	0.00192561980655431\\
	 15	0.389479814677647	0.00795285280371727	0.00372477242513464\\
	 16	0.396229271731051	0.00333334141890163	0.00299329042655289\\
	 17	0.407337205240612	0.00460055593329045	0.00834214116120358\\
	 18	0.414050507718662	0.00591415345735113	0.00414373921006145\\
	 19	0.411884728201864	0.0131203140736862	0.0112265653083287\\
	 20	0.411378998884227	0.0162488576645607	0.0136132231167237\\
	 21	0.415046194615913	0.0173642954889421	0.00601897321432088\\
	 22	0.411477764841811	0.0239940590911523	0.0027555800115755\\
	 23	0.40860618525313	4.45968412759457e-05	6.14133284290164e-05\\
	 24	0.426484292542958	0.0193966659844095	0.0180781254347454\\
	 25	0.412390953779921	0.0360648273709667	0.00402009968383443\\
	 26	0.456183435388939	0.0215091397012018	0.00884577338664844\\
	 27	0.476166197320677	0.00490243139527136	0.00195259449148383\\
	 28	0.475147444653626	0.00852275396741842	0.00488707626396712\\
	 29	0.468883119280803	0.00255850333142349	0.0215098694652565\\
	 30	0.447387728369411	1.94807464092284e-05	4.40523149355387e-05\\
	 31	0.512688810406878	0.0193635291023446	0.00751264968862086\\
	 32	0.514578245810579	0.00465427699315102	0.00331331062199369\\
	 33	0.518936551030115	0.00223097615330903	0.00388599660232425\\
	 34	0.524888156703579	0.00415631915615866	0.0016491820885226\\
	 35	0.531837609018332	0.00579208960065236	0.00405388221162595\\
	 36	0.534641018014775	0.00282393354279065	0.00331505464648973\\
	 37	0.539093663684763	0.00289818753325932	0.0038189675851269\\
	 38	0.543308756760635	0.00292518058421998	0.00830975786945476\\
	 39	0.54777755065673	0.00277408508110399	0.00292023651718398\\
	 40	0.548666715456238	0.005667195457985	0.0166725484053079\\
	 41	0.554741389614051	0.00340843470540764	0.00790520062577393\\
	 42	0.558347903131408	0.00367908050657861	0.00203124553480805\\
	 43	0.561826727866063	0.00459417023232023	0.00352506248292961\\
	 44	0.557203646653997	0.00976697664678583	0.0174403559898791\\
	 45	0.568054069673255	0.00174926392116603	0.00188715665129247\\
	 46	0.570488435426218	0.00424930296451043	0.00310579412815137\\
	 47	0.573794958665764	0.00257045221111341	0.00292853143634453\\
	 48	0.576935646122637	0.00226090742671092	0.0024872865470712\\
	 49	0.579359480089414	0.00176603848700441	0.00334298708767999\\
	 50	0.580868985895391	0.00281888606000602	0.00173642361240034\\
	 51	0.583278759140956	0.00348458051490264	0.00221883780295351\\
	 52	0.586867318995403	0.00347488753394054	0.00348123729499283\\
	 53	0.588917032577715	0.00305997391899016	0.00146765661386883\\
	 54	0.590077976444194	0.00224403896616465	0.00274092671736548\\
	 55	0.591352372863964	0.00368296094309029	0.00281020583789826\\
	 56	0.593712357369033	0.00313709660968553	0.00321107313093849\\
	 57	0.595417598907835	0.00351268302950036	0.0022250292931062\\
	};
	\end{axis}
	\end{tikzpicture}
	\end{small}
	\label{fig:Dim5}
}

\vspace{\fill}

\subfloat[$N=6$]
{
	\begin{small}
	\setlength\figurewidth{6.0cm}
	\setlength\figureheight{6.0cm}
	\begin{tikzpicture}
	
	\begin{axis}[
	width=\figurewidth,
	height=\figureheight,
	scale only axis,
	xmin=6,
	xmax=59,
	xlabel={$M$},
	xmajorgrids,
	ymajorgrids,
	ymin=0.15,
	ymax=0.55,
	ylabel={$\mu(\boldsymbol{A})$},
	legend style={at={(0.97,0.03)},anchor=south east,draw=black,fill=white,legend cell align=left}
	]

	\addplot [
	color=green!60!black,
	only marks,
	thick,
	mark=asterisk,
	mark options={solid}]
	  table[row sep=crcr]{7	0.166666755419515\\
	  8	0.224657895457142\\
	  9	0.250002599554365\\
	  10	0.272254592399196\\
	  11	0.288680915773776\\
	  12	0.3015135433526\\
	  13	0.311924029119596\\
	  14	0.320490883977105\\
	  15	0.327854188906627\\
	  16	0.333375619451399\\
	  17	0.340927213192175\\
	  18	0.346275070978277\\
	  19	0.351458730009994\\
	  20	0.35500334055798\\
	  21	0.357674251420312\\
	  22	0.3607031084508\\
	  23	0.367350396892025\\
	  24	0.372392822643903\\
	  25	0.375911777612163\\
	  26	0.379404757636754\\
	  27	0.382837627387323\\
	  28	0.378870496955219\\
	  29	0.378647151275434\\
	  30	0.372990967791032\\
	  31	0.378556353911323\\
	  32	0.378508931971832\\
	  33	0.378397125021821\\
	  34	0.378286029679276\\
	  35	0.378106282519282\\
	  36	0.377964612173715\\
	  37	0.413824367219636\\
	  38	0.418258967696581\\
	  39	0.424156643998371\\
	  40	0.430261919367656\\
	  41	0.433189153242357\\
	  42	0.439971776862953\\
	  43	0.443546718473432\\
	  44	0.44872552634164\\
	  45	0.453676998658101\\
	  46	0.45734724627255\\
	  47	0.460676862145249\\
	  48	0.464089509404364\\
	  49	0.467255561574741\\
	  50	0.470746041501755\\
	  51	0.47365321915539\\
	  52	0.475858569926095\\
	  53	0.479182088310203\\
	  54	0.481411701629565\\
	  55	0.484240203712435\\
	  56	0.486691810271679\\
	  57	0.488805961288948\\
	  58	0.491325500439037\\
	  59	0.493478479989242\\
	};
	\addlegendentry{BCASC search};
	
	\addplot [
	color=orange!80!black,
	only marks,
	thick,
	mark=o,
	mark options={solid}
	]
	  table[row sep=crcr]{7	0.166815333726216\\
	  8	0.224251752233607\\
	  9	0.250564821265346\\
	  10	0.272957780901266\\
	  11	0.289438235047529\\
	  12	0.302287273542138\\
	  13	0.313482833546503\\
	  14	0.321793591792333\\
	  15	0.328895585712514\\
	  16	0.335993827682243\\
	  17	0.342868428869933\\
	  18	0.350683358742513\\
	  19	0.354835903936877\\
	  20	0.361269112614485\\
	  21	0.362927398682532\\
	  22	0.367634115081585\\
	  23	0.37308276751348\\
	  24	0.385054242179212\\
	  25	0.391499553949207\\
	  26	0.379682165526737\\
	  27	0.379552854599835\\
	  28	0.379057025391604\\
	  29	0.378930872741633\\
	  30	0.378722550818304\\
	  31	0.378508900588839\\
	  32	0.378382025544578\\
	  33	0.378259405957707\\
	  34	0.378164656674124\\
	  35	0.378105015148234\\
	  36	0.378106561072231\\
	  37	0.439649562234088\\
	  38	0.439662076529997\\
	  39	0.447552125271611\\
	  40	0.455610753299527\\
	  41	0.458512458832846\\
	  42	0.463653109637985\\
	  43	0.468188606115307\\
	  44	0.475227404373053\\
	  45	0.476660831275984\\
	  46	0.480766522772093\\
	  47	0.486163310823772\\
	  48	0.489565748655157\\
	  49	0.494151436143536\\
	  50	0.496329726562987\\
	  51	0.498714639815512\\
	  52	0.502125729398265\\
	  53	0.504813165540938\\
	  54	0.5068827461744\\
	  55	0.511730912337755\\
	  56	0.511308722083402\\
	  57	0.515436274408078\\
	  58	0.517356438544456\\
	  59	0.52154451840023\\
	};
	\addlegendentry{Approach by~\cite{medra_flexible_2014}};
	
	\addplot [
	color=blue,
	thick,
	densely dotted]
	table[row sep=crcr]{7	0.166666666666667\\
	8	0.218217890235992\\
	9	0.25\\
	10	0.272165526975909\\
	11	0.288675134594813\\
	12	0.301511344577764\\
	13	0.311804782231162\\
	14	0.320256307610174\\
	15	0.327326835353989\\
	16	0.333333333333333\\
	17	0.338501600193165\\
	18	0.342997170285018\\
	19	0.346944333244355\\
	20	0.350438322025231\\
	21	0.353553390593274\\
	22	0.356348322549899\\
	23	0.358870281282637\\
	24	0.361157559257308\\
	25	0.363241578628389\\
	26	0.365148371670111\\
	27	0.366899692852671\\
	28	0.368513865595044\\
	29	0.370006434950477\\
	30	0.371390676354104\\
	31	0.372677996249965\\
	32	0.373878250552983\\
	33	0.375\\
	34	0.376050716545177\\
	35	0.377036951431949\\
	36	0.377964473009227\\
	37	0.408248290463863\\
	38	0.408248290463863\\
	39	0.408248290463863\\
	40	0.408248290463863\\
	41	0.408248290463863\\
	42	0.408248290463863\\
	43	0.412170073979383\\
	44	0.415851332602251\\
	45	0.419313934688767\\
	46	0.422577127364258\\
	47	0.425657924852277\\
	48	0.428571428571429\\
	49	0.431331092813754\\
	50	0.433948946665029\\
	51	0.436435780471985\\
	52	0.438801303383356\\
	53	0.441054277135977\\
	54	0.443202630213959\\
	55	0.445253555699417\\
	56	0.447213595499958\\
	57	0.449088713139072\\
	58	0.450884356899531\\
	59	0.452605514793582\\
	};
	\addlegendentry{Lower bound~\eqref{eq:LowerBoundWhereWelchBoundFails}};
	
	\addplot [
	color=blue,
	thick,
	solid]
	table[row sep=crcr]{7	0.166666666666667\\
	8	0.218217890235992\\
	9	0.25\\
	10	0.272165526975909\\
	11	0.288675134594813\\
	12	0.301511344577764\\
	13	0.311804782231162\\
	14	0.320256307610174\\
	15	0.327326835353989\\
	16	0.333333333333333\\
	17	0.338501600193165\\
	18	0.342997170285018\\
	19	0.346944333244355\\
	20	0.350438322025231\\
	21	0.353553390593274\\
	22	0.356348322549899\\
	23	0.358870281282637\\
	24	0.361157559257308\\
	25	0.363241578628389\\
	26	0.365148371670111\\
	27	0.366899692852671\\
	28	0.368513865595044\\
	29	0.370006434950477\\
	30	0.371390676354104\\
	31	0.372677996249965\\
	32	0.373878250552983\\
	33	0.375\\
	34	0.376050716545177\\
	35	0.377036951431949\\
	36	0.377964473009227\\
	};
	
	\addplot [color=green!80!black,only marks,mark=-,mark options={solid},forget plot,thick]
	 plot [error bars/.cd, y dir = both, y explicit, error mark options={rotate=90, thick}, error bar style={thick}]
	 table[row sep=crcr, y error plus index=2, y error minus index=3]{7	0.166666805825078	7.70809614747403e-08	5.04055634453771e-08\\
	 8	0.224657911359925	1.08445490365483e-08	1.59027828128266e-08\\
	 9	0.250005794499142	8.58615885446223e-07	3.19494477701854e-06\\
	 10	0.272254997204992	1.35834506737043e-06	4.04805796827556e-07\\
	 11	0.288718002247728	9.46563365022968e-06	3.70864739516752e-05\\
	 12	0.301533193466398	4.66274034464087e-05	1.96501137970095e-05\\
	 13	0.312111094770951	0.000121714457750666	0.000187065651354845\\
	 14	0.320676833062333	0.000111956474095176	0.000185949085228843\\
	 15	0.32821733639899	0.000474997222745188	0.00036314749236277\\
	 16	0.334278877926735	0.00134038119578833	0.000903258475335955\\
	 17	0.341454663680507	0.000788154944697084	0.000527450488332248\\
	 18	0.347253721725132	0.000949313783054462	0.000978650746855048\\
	 19	0.352723464559974	0.00101754766424633	0.00126473454998027\\
	 20	0.356009090464419	0.00220944793956407	0.0010057499064387\\
	 21	0.358777693247553	0.00169637647674098	0.0011034418272412\\
	 22	0.362405603889164	0.00264307930551738	0.00170249543836409\\
	 23	0.368614001845111	0.00165539066359943	0.00126360495308575\\
	 24	0.373066298382127	0.000738666095606633	0.000673475738224072\\
	 25	0.377334727975202	0.00117891250441843	0.00142295036303869\\
	 26	0.380770209305767	0.000730679157342706	0.00136545166901286\\
	 27	0.38392934631872	0.000993774582201234	0.00109171893139692\\
	 28	0.38473733621042	0.00318636481943563	0.00586683925520154\\
	 29	0.38759482120568	0.00315025249446282	0.0089476699302462\\
	 30	0.387218328056828	0.00644731467522824	0.0142273602657964\\
	 31	0.386715103859152	0.00849160505551672	0.00815874994782861\\
	 32	0.382327713120266	0.0151172652366449	0.00381878114843398\\
	 33	0.380482912040494	0.0182642464565157	0.00208578701867312\\
	 34	0.379881410610859	0.0141623808213961	0.00159538093158307\\
	 35	0.378106887085801	2.95428136831699e-06	6.0456651912455e-07\\
	 36	0.380485130765198	0.022677560825454	0.00252051859148239\\
	 37	0.414232907688636	0.00080868979445009	0.000408540468999852\\
	 38	0.419637417996549	0.00202442864336766	0.00137845029996841\\
	 39	0.425996147469628	0.00125976960264362	0.00183950347125639\\
	 40	0.43105771940553	0.00119489623496682	0.000795800037874272\\
	 41	0.435814455246207	0.00108767618155281	0.00262530200384981\\
	 42	0.440976376886664	0.000834363296367879	0.00100460002371122\\
	 43	0.445516112716458	0.00128894591472339	0.00196939424302522\\
	 44	0.449551457118427	0.000575753778731647	0.00082593077678661\\
	 45	0.454220809582553	0.000572577533478835	0.000543810924452492\\
	 46	0.457955983829131	0.000491726368887224	0.000608737556581029\\
	 47	0.461406956825089	0.000655754689591947	0.000730094679840476\\
	 48	0.464870122441513	0.000705097428973012	0.00078061303714877\\
	 49	0.467785527949861	0.000457748340099651	0.000529966375120372\\
	 50	0.471012320723351	0.000522047851029772	0.000266279221596\\
	 51	0.474123351970174	0.00037497539871284	0.000470132814784474\\
	 52	0.476601782770984	0.000571395475687464	0.000743212844889152\\
	 53	0.479710253870366	0.000553870219591801	0.000528165560162719\\
	 54	0.482042518887204	0.000429324497823125	0.000630817257639249\\
	 55	0.484659703248948	0.000551767156463656	0.000419499536513324\\
	 56	0.487047183541905	0.000519017654027398	0.000355373270226167\\
	 57	0.489433795732024	0.000682212379024161	0.00062783444307607\\
	 58	0.491854683737676	0.000607849532465898	0.000529183298638991\\
	 59	0.493910661424511	0.000930328701756966	0.000432181435268741\\
	};
	\addplot [color=orange!80!white,only marks,mark=-,mark options={solid},forget plot, thick]
	 plot [error bars/.cd, y dir = both, y explicit, error mark options={rotate=90, thick}, error bar style={thick}]
	 table[row sep=crcr, y error plus index=2, y error minus index=3]{7	0.166894273862368	7.65962521510344e-05	7.89401361521447e-05\\
	 8	0.224291818923097	7.49305297364222e-05	4.00666894898039e-05\\
	 9	0.250670687582797	0.000153145585258863	0.000105866317451353\\
	 10	0.273288050869293	0.00026934038301335	0.000330269968027264\\
	 11	0.290891195925632	0.00140154578299884	0.00145296087810276\\
	 12	0.303125437005223	0.00161130560245382	0.00083816346308474\\
	 13	0.314142095804946	0.00127180722669917	0.000659262258442705\\
	 14	0.322727584377396	0.00100895378684229	0.000933992585062493\\
	 15	0.33035924569106	0.00135106202678498	0.00146365997854508\\
	 16	0.338091190866164	0.0017415422576299	0.0020973631839209\\
	 17	0.345721376167784	0.00233892253937962	0.00285294729785096\\
	 18	0.352280965504579	0.00246236146641854	0.00159760676206622\\
	 19	0.356993668943454	0.00215587744977586	0.00215776500657622\\
	 20	0.364115760178527	0.0040963151262482	0.00284664756404202\\
	 21	0.367141537003974	0.00668893035389079	0.00421413832144146\\
	 22	0.37540274748399	0.00469397491978929	0.00776863240240494\\
	 23	0.381525192613214	0.00605048778313605	0.00844242509973403\\
	 24	0.389071698070366	0.00322189765449027	0.00401745589115393\\
	 25	0.395122862173672	0.00240884615224829	0.00362330822446533\\
	 26	0.39726563512685	0.00493523187901218	0.0175834696001131\\
	 27	0.398496330107341	0.00760855369186575	0.0189434755075055\\
	 28	0.400461291576347	0.00825012144312537	0.0214042661847436\\
	 29	0.399798567311023	0.0102985460971999	0.0208676945693905\\
	 30	0.395524790620174	0.0201998579427147	0.0168022398018696\\
	 31	0.401692271707482	0.0201865795262495	0.0231833711186425\\
	 32	0.40412702685814	0.0210978877808644	0.0257450013135616\\
	 33	0.40986705501712	0.0169277389336426	0.031607649059413\\
	 34	0.395211232932796	0.0325591588743463	0.0170465762586721\\
	 35	0.414824806339568	0.019513446709238	0.0367197911913341\\
	 36	0.426056793301175	0.0115989791817593	0.0479502322289441\\
	 37	0.447227818582255	0.019163611681948	0.0075782563481665\\
	 38	0.446474795208638	0.0056632123072457	0.00681271867864097\\
	 39	0.45192977978794	0.00827353419252602	0.00437765451632877\\
	 40	0.457866526630967	0.00324101199517235	0.00225577333143989\\
	 41	0.46168971459	0.00479088409552869	0.00317725575715466\\
	 42	0.46691316493913	0.002539905931776	0.0032600553011447\\
	 43	0.471569962102731	0.00306721818141792	0.00338135598742412\\
	 44	0.477477474851226	0.0018855203334282	0.00225007047817294\\
	 45	0.480830110851161	0.00369066903381854	0.00416927957517765\\
	 46	0.484049049521253	0.00153465174743894	0.00328252674916008\\
	 47	0.488115229888766	0.00389871969529249	0.00195191906499453\\
	 48	0.492214274681219	0.00260330284208249	0.00264852602606203\\
	 49	0.495332992905023	0.00219513599512483	0.00118155676148662\\
	 50	0.498044717637429	0.00238199251839288	0.00171499107444206\\
	 51	0.500918046565732	0.00215927815010541	0.00220340675021996\\
	 52	0.504742254849724	0.00361262756213376	0.00261652545145885\\
	 53	0.507611612665225	0.00383618004745634	0.00279844712428634\\
	 54	0.508950926229801	0.00197324072834892	0.00206818005540077\\
	 55	0.513013360221818	0.0052825243656035	0.00128244788406351\\
	 56	0.5150574963306	0.00474827948955003	0.00374877424719855\\
	 57	0.517430532613231	0.00520860498688858	0.00199425820515375\\
	 58	0.519218722266135	0.0015537989177471	0.00186228372167851\\
	 59	0.522616353396356	0.00141288047462429	0.0010718349961264\\
	};
	\end{axis}
	\end{tikzpicture}
	\end{small}
	\label{fig:Dim6}
}
\hfil
\subfloat[$N=7$]
{
	\begin{small}
	\setlength\figurewidth{6.0cm}
	\setlength\figureheight{6.0cm}
	\begin{tikzpicture}
	
	\begin{axis}[
	width=\figurewidth,
	height=\figureheight,
	scale only axis,
	xmin=8,
	xmax=53,
	xlabel={$M$},
	xmajorgrids,
	ymajorgrids,
	ymin=0.1,
	ymax=0.5,
	ylabel={$\mu(\boldsymbol{A})$},
	legend style={at={(0.97,0.03)},anchor=south east,draw=black,fill=white,legend cell align=left}
	]

	\addplot [
	color=green!60!black,
	only marks,
	thick,
	mark=asterisk,
	mark options={solid}]
	  table[row sep=crcr]{8	0.142857231997501\\
	  9	0.196697542643602\\
	  10	0.221761926890348\\
	  11	0.239476088844277\\
	  12	0.255333444382798\\
	  13	0.267321029418286\\
	  14	0.277359577400295\\
	  15	0.285807064306452\\
	  16	0.292899490988721\\
	  17	0.299093786035107\\
	  18	0.304468525865639\\
	  19	0.309383849829285\\
	  20	0.314381845638088\\
	  21	0.318746702357028\\
	  22	0.323180409129437\\
	  23	0.327182694098801\\
	  24	0.330649133146914\\
	  25	0.334751205961361\\
	  26	0.334828903295814\\
	  27	0.334992304960719\\
	  28	0.333377249881184\\
	  29	0.344120339973182\\
	  30	0.3477532413819\\
	  31	0.351203965538266\\
	  32	0.354117566830368\\
	  33	0.356997927066685\\
	  34	0.359367590429811\\
	  35	0.362120940355052\\
	  36	0.364052476893528\\
	  37	0.365586103340457\\
	  38	0.368224257362774\\
	  39	0.370117655264858\\
	  40	0.371282191657991\\
	  41	0.370515156109253\\
	  42	0.354280283716146\\
	  43	0.354267103120534\\
	  44	0.354145629343466\\
	  45	0.354013005038757\\
	  46	0.353933338972087\\
	  47	0.353815524914474\\
	  48	0.353668640610907\\
	  49	0.353553930853742\\
	  50	0.388704110800823\\
	  51	0.394669316631092\\
	  52	0.39958769961533\\
	  53	0.40330051315738\\
	};
	\addlegendentry{BCASC search};
	
	\addplot [
	color=orange!80!black,
	only marks,
	thick,
	mark=o,
	mark options={solid}
	]
	  table[row sep=crcr]{8	0.143027082553256\\
	  9	0.196265405351048\\
	  10	0.2217920137376\\
	  11	0.239723640429893\\
	  12	0.256397241517155\\
	  13	0.268489121298423\\
	  14	0.27834945340223\\
	  15	0.287069595458846\\
	  16	0.294452793133975\\
	  17	0.300367761390126\\
	  18	0.306240992447834\\
	  19	0.311270989269595\\
	  20	0.318167320527088\\
	  21	0.321920585311707\\
	  22	0.328750269248724\\
	  23	0.331834530824183\\
	  24	0.336979402547172\\
	  25	0.335530038531942\\
	  26	0.335243426099521\\
	  27	0.351407046229427\\
	  28	0.334616030817777\\
	  29	0.359593257067193\\
	  30	0.363658661593963\\
	  31	0.366687037227921\\
	  32	0.371710103621185\\
	  33	0.375523661619389\\
	  34	0.379075175281733\\
	  35	0.383444726141212\\
	  36	0.38560264927638\\
	  37	0.388777582089147\\
	  38	0.388138923439923\\
	  39	0.395036865411365\\
	  40	0.395951558907358\\
	  41	0.398645862036864\\
	  42	0.400275014364986\\
	  43	0.403789293165798\\
	  44	0.40481262983394\\
	  45	0.407839133202475\\
	  46	0.353764420566738\\
	  47	0.411535252416479\\
	  48	0.353685819905826\\
	  49	0.415931103560753\\
	  50	0.413542114015042\\
	  51	0.423495026175661\\
	  52	0.426665284951799\\
	  53	0.429776215620876\\
	};
	\addlegendentry{Approach by~\cite{medra_flexible_2014}};
	
	\addplot [
	color=blue,
	thick,
	densely dotted]
	table[row sep=crcr]{8	0.142857142857143\\
	9	0.188982236504614\\
	10	0.218217890235992\\
	11	0.239045721866879\\
	12	0.254823595718813\\
	13	0.267261241912424\\
	14	0.277350098112615\\
	15	0.285714285714286\\
	16	0.29277002188456\\
	17	0.298807152333598\\
	18	0.30403449605253\\
	19	0.308606699924184\\
	20	0.312640945658523\\
	21	0.316227766016838\\
	22	0.31943828249997\\
	23	0.322329185610152\\
	24	0.324946244957226\\
	25	0.327326835353989\\
	26	0.329501788419166\\
	27	0.331496772065898\\
	28	0.333333333333333\\
	29	0.335029697130245\\
	30	0.336601385337074\\
	31	0.338061701891407\\
	32	0.339422116651065\\
	33	0.340692571934623\\
	34	0.341881729378914\\
	35	0.342997170285018\\
	36	0.344045559394066\\
	37	0.345032779671177\\
	38	0.345964043928151\\
	39	0.346843987809648\\
	40	0.347676747682558\\
	41	0.348466026218585\\
	42	0.349215147884789\\
	43	0.349927106111883\\
	44	0.350604603563426\\
	45	0.351250086657104\\
	46	0.351865775274498\\
	47	0.352453688425121\\
	48	0.3530156664941\\
	49	0.353553390593274\\
	50  0.377964473\\
	51  0.377964473\\
	52  0.377964473\\
	53  0.377964473\\
	};
	\addlegendentry{Lower bound~\eqref{eq:LowerBoundWhereWelchBoundFails}};
	
	\addplot [
	color=blue,
	thick,
	solid]
	table[row sep=crcr]{8	0.142857142857143\\
	9	0.188982236504614\\
	10	0.218217890235992\\
	11	0.239045721866879\\
	12	0.254823595718813\\
	13	0.267261241912424\\
	14	0.277350098112615\\
	15	0.285714285714286\\
	16	0.29277002188456\\
	17	0.298807152333598\\
	18	0.30403449605253\\
	19	0.308606699924184\\
	20	0.312640945658523\\
	21	0.316227766016838\\
	22	0.31943828249997\\
	23	0.322329185610152\\
	24	0.324946244957226\\
	25	0.327326835353989\\
	26	0.329501788419166\\
	27	0.331496772065898\\
	28	0.333333333333333\\
	29	0.335029697130245\\
	30	0.336601385337074\\
	31	0.338061701891407\\
	32	0.339422116651065\\
	33	0.340692571934623\\
	34	0.341881729378914\\
	35	0.342997170285018\\
	36	0.344045559394066\\
	37	0.345032779671177\\
	38	0.345964043928151\\
	39	0.346843987809648\\
	40	0.347676747682558\\
	41	0.348466026218585\\
	42	0.349215147884789\\
	43	0.349927106111883\\
	44	0.350604603563426\\
	};
	
	\addplot [color=green!80!black,only marks,mark=-,mark options={solid},forget plot,thick]
	 plot [error bars/.cd, y dir = both, y explicit, error mark options={rotate=90, thick}, error bar style={thick}]
	 table[row sep=crcr, y error plus index=2, y error minus index=3]{8	0.14285730678971	9.49370283220841e-08	7.47922092592113e-08\\
	 9	0.196698082385238	1.73398301375749e-06	5.39741636107482e-07\\
	 10	0.221762341560578	4.29690469455624e-07	4.14670230125536e-07\\
	 11	0.239494436847697	2.76233876491672e-05	1.83480034199879e-05\\
	 12	0.255375129193067	0.000159883388888671	4.16848102686718e-05\\
	 13	0.267419060555931	9.54341002380454e-05	9.80311376455045e-05\\
	 14	0.277469266437869	0.000159916948572592	0.000109689037573746\\
	 15	0.2859473123168	0.000260663889596724	0.000140248010348121\\
	 16	0.293070070863478	0.000200740558278956	0.00017057987475616\\
	 17	0.299263827039	0.000155959828159247	0.000170041003892896\\
	 18	0.304780324777485	0.000637845747347932	0.000311798911846517\\
	 19	0.309885309234991	0.00043045878887854	0.000501459405706062\\
	 20	0.314931903260164	0.00031713155627705	0.000550057622075828\\
	 21	0.319577672363143	0.000704948675461947	0.000830970006115206\\
	 22	0.323670084294708	0.000463484991311436	0.000489675165271319\\
	 23	0.327861584275156	0.000520996531857054	0.000678890176355751\\
	 24	0.331483072593629	0.00121588258417032	0.000833939446714838\\
	 25	0.335265008094051	0.00060559724800735	0.000513802132690311\\
	 26	0.337795988971906	0.00142124885873479	0.00296708567609211\\
	 27	0.339373031188823	0.00234782591769556	0.0043807262281042\\
	 28	0.338819846203815	0.00646741065843104	0.00544259632263167\\
	 29	0.346151069488173	0.00122665601803768	0.0020307295149905\\
	 30	0.348973332630986	0.00110506409818151	0.00122009124908556\\
	 31	0.352390248794345	0.000984663272887321	0.0011862832560785\\
	 32	0.354694440070248	0.000697353945158408	0.000576873239880038\\
	 33	0.357509111743349	0.000576968081033524	0.000511184676663612\\
	 34	0.360137459524537	0.000960428499507782	0.000769869094726205\\
	 35	0.362635870874868	0.000553281623589807	0.000514930519816381\\
	 36	0.364794427351655	0.000745777822124705	0.000741950458127194\\
	 37	0.366558447606337	0.000610896970148878	0.000972344265879943\\
	 38	0.36867963124362	0.000549275927223425	0.000455373880845433\\
	 39	0.370643616697338	0.000735634488431003	0.000525961432480182\\
	 40	0.372275981437297	0.000810401654973647	0.000993789779306309\\
	 41	0.373710320309294	0.00121316268040755	0.00319516420004096\\
	 42	0.373201742116088	0.00343744078677388	0.0189214583999419\\
	 43	0.37391179363694	0.00358378638212464	0.0196446905164065\\
	 44	0.375338597138536	0.00422713303133138	0.0211929677950695\\
	 45	0.371761758764525	0.0090181175021159	0.0177487537257682\\
	 46	0.367230149859816	0.014313365260242	0.0132968108877287\\
	 47	0.361565641393314	0.0207942561964203	0.00775011647884011\\
	 48	0.380245298583629	0.00405575849416784	0.026576657972722\\
	 49	0.375194674500874	0.0102423828004216	0.0216407436471321\\
	 50	0.390781147910526	0.00136239140923977	0.00207703710970319\\
	 51	0.39579461382893	0.000694663224270475	0.0011252971978386\\
	 52	0.40010955085999	0.000487558870446381	0.000521851244660254\\
	 53	0.403964725275714	0.000546289586967841	0.000664212118334062\\
	};
	\addplot [color=orange!80!white,only marks,mark=-,mark options={solid},forget plot, thick]
	 plot [error bars/.cd, y dir = both, y explicit, error mark options={rotate=90, thick}, error bar style={thick}]
	 table[row sep=crcr, y error plus index=2, y error minus index=3]{8	0.143081980589311	5.96899371196891e-05	5.489803605499e-05\\
	 9	0.196386494018153	0.000160667484932431	0.00012108866710514\\
	 10	0.221944532583456	0.000166070437236876	0.000152518845856348\\
	 11	0.240600539587849	0.00081231371804541	0.000876899157955635\\
	 12	0.256917057161948	0.000406003736676486	0.000519815644792254\\
	 13	0.269375474177275	0.0007728945288073	0.000886352878851748\\
	 14	0.278990104707833	0.000448846021643967	0.000640651305602424\\
	 15	0.287447031944468	0.000580334399696569	0.000377436485622051\\
	 16	0.29515163804197	0.000723273021442206	0.000698844907995488\\
	 17	0.301196230677709	0.00106419654426942	0.000828469287583289\\
	 18	0.307210339559316	0.000631428752079832	0.000969347111481833\\
	 19	0.312980542107873	0.00147396975912878	0.00170955283827801\\
	 20	0.319245191947669	0.00164341899476711	0.0010778714205813\\
	 21	0.324391620371833	0.00215483470475014	0.00247103506012586\\
	 22	0.330173234802228	0.00289125006148933	0.00142296555350435\\
	 23	0.334839981799839	0.00286230232202322	0.00300545097565647\\
	 24	0.339650032074247	0.00184025462421145	0.00267062952707509\\
	 25	0.342339814037363	0.00446993549395708	0.00680977550542156\\
	 26	0.344479948671888	0.00467446076867334	0.00923652257236734\\
	 27	0.353486080090974	0.00220078874112928	0.00207903386154717\\
	 28	0.355961415928762	0.00365799791605947	0.021345385110985\\
	 29	0.361708537136872	0.00303598630731533	0.00211528006967943\\
	 30	0.365499089476521	0.00215342521150624	0.00184042788255745\\
	 31	0.370730422373127	0.00625377278716482	0.00404338514520619\\
	 32	0.373808115867007	0.00156484332672696	0.00209801224582268\\
	 33	0.378056008173729	0.00415446524878399	0.00253234655433948\\
	 34	0.381269648468477	0.00127374323685503	0.00219447318674426\\
	 35	0.38527385227505	0.00269325472331067	0.00182912613383851\\
	 36	0.387992633217699	0.00173097885454171	0.00238998394131867\\
	 37	0.39101806581407	0.00198768573179547	0.00224048372492225\\
	 38	0.391297944251204	0.00240280246897823	0.00315902081128167\\
	 39	0.396448344233911	0.00172099654309815	0.00141147882254589\\
	 40	0.39875216524587	0.00443159446650288	0.00280060633851192\\
	 41	0.402008177799122	0.00288606850475831	0.00336231576225776\\
	 42	0.40254045090553	0.00366096645561331	0.00226543654054429\\
	 43	0.405988911594263	0.00162048060359066	0.00219961842846422\\
	 44	0.408813916979723	0.0039135138114792	0.00400128714578235\\
	 45	0.411216974492092	0.00373766975564871	0.00337784128961693\\
	 46	0.402082055976916	0.0155698240554392	0.0483176354101785\\
	 47	0.416535438718752	0.00213520032290893	0.00500018630227306\\
	 48	0.414235954668895	0.0100289009571045	0.0605501347630695\\
	 49	0.421153951371855	0.00312655974203585	0.00522284781110227\\
	 50	0.422423441960946	0.0282912468627727	0.00888132794590396\\
	 51	0.425352711032546	0.00326293750967871	0.0018576848568852\\
	 52	0.428634716615922	0.00152740167157739	0.00196943166412278\\
	 53	0.432059117423156	0.002100462003577	0.00228290180227997\\
	};
	\end{axis}
	\end{tikzpicture}%
	\end{small}
	\label{fig:Dim7}
}
\caption{\label{fig:Dim4-7}Best obtained coherence out of ten runs for each number of vectors~$M$ in $N$ dimensions. The proposed BCASC search method is compared to the approach in~\cite{medra_flexible_2014}. The errorbars indicate the variance in the ten runs. The line for the lower bound is solid if~\eqref{eq:WelchLimits} is fulfilled.}
\end{figure*}
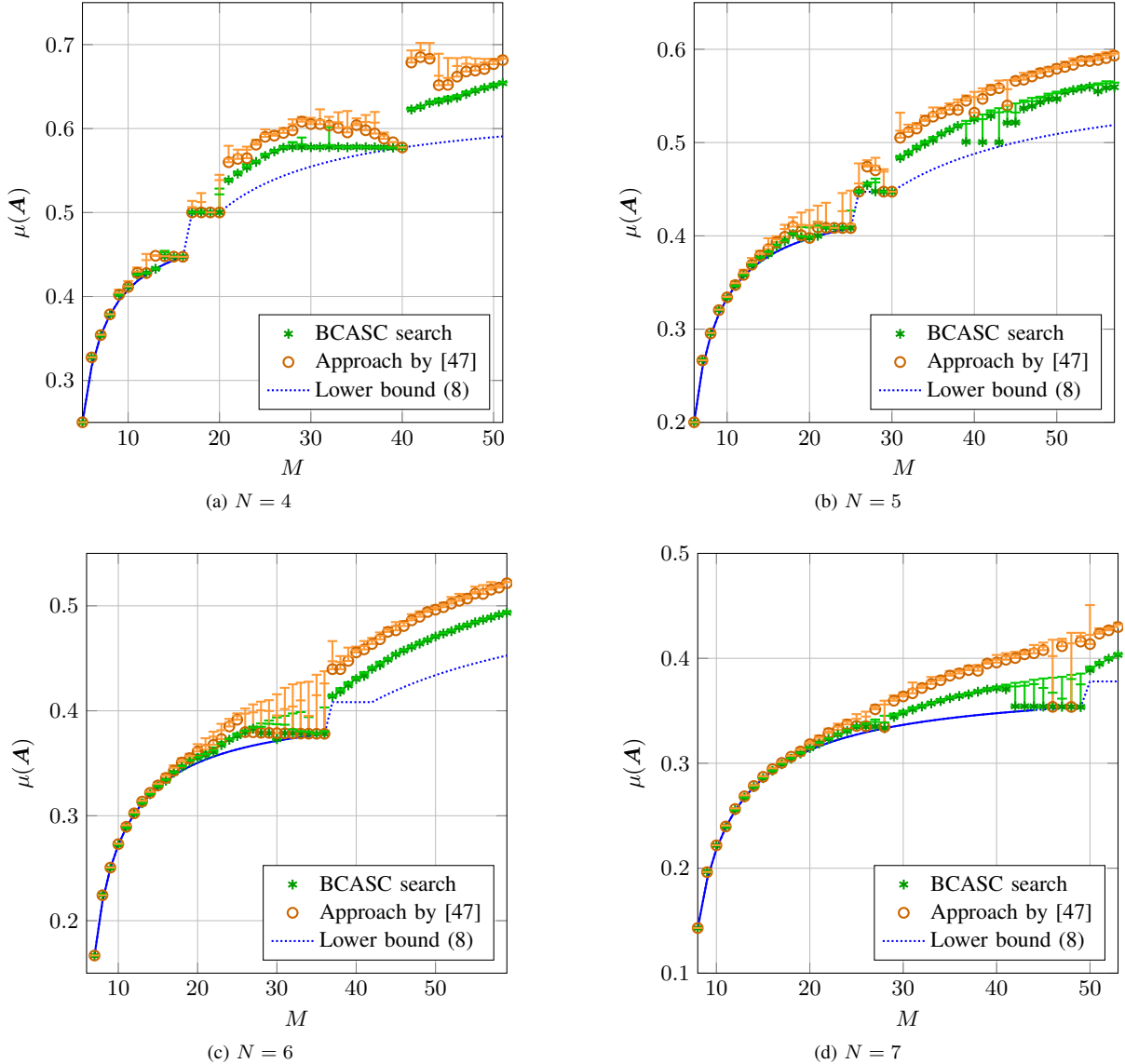
The results for the dimensions $N=4,\hdots,7$ are displayed in Fig.~\ref{fig:Dim4-7}.
The coherence values obtained by both methods are almost equivalent for $M\leq20,30,36\text{, and }28$ respectively.
In Fig.~\ref{fig:Dim4} and \ref{fig:Dim5}, there is again a level of constant coherence observable for $M>N^2$ after which our method always yields smaller coherence values than the approach of~\cite{medra_flexible_2014}.
This plateau corresponds to the orthoplex bound~\eqref{eq:orthoplex}, and its length equals the individual dimensionality~$N$ for the mentioned cases, since it is limited by~\eqref{eq:Levenshtein}.
Such an almost constant level can also be observed for $N=2$ (cf.~\cite[Fig.~3]{xia_achieving_2005}).
Each obtained vector set with $M=N^2+N$ and $N=3,\hdots,5$ can be sorted into a matrix~$\boldsymbol{A}$ such that the corresponding Gram matrix $\boldsymbol{G}=\boldsymbol{A}^{\mathrm{H}}\boldsymbol{A}$ shows a block diagonal structure:
There are $N+1$ identity matrices of dimension $N$ on the diagonal while all other blocks contain entries of constant modulus.
The vectors of each diagonal block correspond to an $N$-dimensional subspace. 
Thus, the obtained solutions can be interpreted as Grassmannian subspace packings $\mathcal{G}(N,N)$ of $N+1$ subspaces.
These vector sets correspond to the previously mentioned constructions of MUBs given in~\cite{wootters_optimal_1989,ding_signal_2007}.
For $N=6$ and~$7$ in Fig.~\ref{fig:Dim6} and~\ref{fig:Dim7}, such a quasi-constant level has not been found by the described methods.
Since $N=6$ is not a prime power, it is already questioned in~\cite{wootters_optimal_1989} whether a vector set of $M=42$ vectors exists which achieves equality in~\eqref{eq:orthoplex} and~\eqref{eq:Levenshtein}. 
Other plateaus can also be found implicitly:
In such a case, a vector set with smaller coherence but larger cardinality $M$ is obtained by optimization. As consequence, vectors can be deleted from the found set while the low coherence remains the same.
For example, observe that there is a plateau for $35\leq M\leq43$ implicitly given by $M=43$ in Fig.~\ref{fig:Dim5}.
Plateaus of almost constant coherence also usually precede the optimal constellation of $M=N^2$.
Increasing dimensionality favors our BCASC search approach, since it obtains vector sets with smaller coherence also in the range of $M<N^2$ starting from $M$ slightly larger than $N$. (cf.~Fig.~\ref{fig:Dim6} and~\ref{fig:Dim7}).
Further simulations confirm this trend, for example, our optimized vector sets have smaller coherence for $M\geq12$ with $N=8\text{ and }9$.

In Table~\ref{tab:CoherenceTabXiLove}, we considered also the matrices provided by~\cite{love_index_????} and the results of\cite{xia_achieving_2005} for comparison.
\begin{table}[!t]
\renewcommand{\arraystretch}{1.3}
\centering
\caption{\label{tab:CoherenceTabXiLove}Comparison of Numerical Search Algorithms as in~\cite{xia_achieving_2005}}
\begin{tabular}{r|r|C{1.0cm}|C{1.05cm}|C{1.0cm}|C{1.0cm}|C{1.17cm}}
\hlx{hv}
$N$&$M$& BCASC search& Medra et al. \cite{medra_flexible_2014}& \centerline{Love} \newline \centering \cite{love_index_????}& Xia et al. \cite{xia_achieving_2005}& composite bound \eqref{eq:LowerBoundWhereWelchBoundFails}\\\hlx{vhv}
$2$&$8$&$0.7950$&$0.7997$&$0.8415$&$0.8216$&$0.7500$\\\hlx{vhv}
$3$&$16$&$0.6491$&$0.6590$&$0.8079$&$0.6766$&$0.6202$\\\hlx{vhv}
$4$&$16$&$0.4472$&$0.4473$&$0.7525$&$0.4514$&$0.4472$\\\hlx{vhv}
$4$&$64$&$0.6869$&$0.7151$&$0.7973$&$0.7447$&$0.6000$\\\hlx{vh}
\end{tabular}
\end{table}
It should be noted that, with exception of the case $N=4$, $M=16$, the Welch bound cannot be obtained since $M>N^2$ [cf.~\eqref{eq:WelchLimits}].
The presented BCASC search approach is able to reach the Welch bound for this case, and it yields vector sets with the smallest coherence for the other cases.

Additionally, we compared the obtained results to those of~\cite{dhilon_constructing_2008} with respect to the coherence in Table~\ref{tab:CoherenceTabDhiMe} as it is similarly done in~\cite[Tab. II]{medra_flexible_2014}.
\begin{table}[!t]
\renewcommand{\arraystretch}{1.3}
\centering
\caption{\label{tab:CoherenceTabDhiMe}Comparison of Numerical Search Algorithms as in~\cite{medra_flexible_2014}}
\begin{tabular}{r|r|C{1.0cm}|C{1.05cm}|C{1.1cm}|C{1.17cm}}
\hlx{hv}
$N$&$M$& BCASC search& Medra et al. \cite{medra_flexible_2014}& Dhilon et al. \cite{dhilon_constructing_2008}& composite bound \eqref{eq:LowerBoundWhereWelchBoundFails}\\\hlx{vhv}
$4$&$5$&$0.2500$&$0.2502$&$0.2500$&$0.2500$\\\hlx{vhv}
$4$&$6$&$0.3277$&$0.3274$&$0.3275$&$0.3162$\\\hlx{vhv}
$4$&$7$&$0.3536$&$0.3540$&$0.3536$&$0.3536$\\\hlx{vhv}
$4$&$8$&$0.3780$&$0.3787$&$0.3782$&$0.3780$\\\hlx{vhv}
$4$&$9$&$0.4022$&$0.4021$&$0.4034$&$0.3953$\\\hlx{vhv}
$4$&$10$&$0.4118$&$0.4113$&$0.4114$&$0.4082$\\\hlx{vhv}
$4$&$16$&$0.4472$&$0.4473$&$0.4473$&$0.4472$\\\hlx{vhv}
$4$&$20$&$0.5000$&$0.5001$&$0.5335$&$0.5000$\\\hlx{vhv}
$5$&$6$&$0.2000$&$0.2002$&$0.2001$&$0.2000$\\\hlx{vhv}
$5$&$7$&$0.2670$&$0.2665$&$0.2669$&$0.2582$\\\hlx{vhv}
$5$&$8$&$0.2955$&$0.2954$&$0.2955$&$0.2928$\\\hlx{vhv}
$5$&$9$&$0.3207$&$0.3203$&$0.3216$&$0.3162$\\\hlx{vhv}
$5$&$10$&$0.3333$&$0.3341$&$0.3336$&$0.3333$\\\hlx{vhv}
$5$&$16$&$0.3889$&$0.3932$&$0.3959$&$0.3830$\\\hlx{vh}
\end{tabular}
\end{table}
It can be seen from the results that our approach reaches the composite bound~\eqref{eq:LowerBoundWhereWelchBoundFails} most often.
In cases where the bound could not be reached, we obtained slightly better results with the approach of~\cite{medra_flexible_2014}.
The algorithm of~\cite{dhilon_constructing_2008} shows in general the worst performance within this comparison.
However, with exception of $N=4$ and $M=20$, there are no significant negative outliers for the investigated range of $N$ and $M$.

As drawback, it needs to be mentioned that the numerical integration in the proposed BCASC search approach is, especially for large combinations of $N$ and $M$, computational costly.
Therefore, the choice of the numerical search algorithm depends on the needed level of coherence and the available computational resources.
The influence of the numerical integration on the performance of the search algorithm is investigated in the following.

\subsection{Evaluation of Integral Approximation}
\label{sec:NumSimIntegralApprox}
\begin{figure}[!t]
\centering
\begin{small}
\setlength\figurewidth{7.0cm}
\setlength\figureheight{7.0cm}
\begin{tikzpicture}

\begin{axis}[
width=\figurewidth,
height=\figureheight,
scale only axis,
xmin=1,
xmax=24,
xlabel={$K$},
xmajorgrids,
ymajorgrids,
ymin=0.43,
ymax=1,
ylabel={$\mu(\boldsymbol{A})$},
extra x ticks={22},
extra x tick style={grid=major, tick label style={anchor=north}},
extra x tick labels={$22$},
legend style={draw=black,fill=white,legend cell align=left}
]
\addplot [color=blue,thick,solid,mark=asterisk,mark options={solid}]
  table[row sep=crcr]{1	1\\
2	0.978686602123639\\
3	0.924427934229156\\
4	0.883566727958792\\
5	0.858337963297972\\
6	0.8063030852508\\
7	0.812705980080014\\
8	0.805978687572681\\
9	0.807988775832918\\
10	0.804432600389542\\
11	0.805248170511662\\
12	0.804332244501687\\
13	0.804229095503741\\
14	0.80052773444613\\
15	0.800907143401005\\
16	0.798853123991185\\
17	0.799499226138601\\
18	0.798614498513003\\
19	0.798118737616691\\
20	0.794935902334084\\
21	0.796958968029686\\
22	0.797076862247301\\
23	0.797145173418235\\
24	0.796964674807591\\
};
\addlegendentry{$N=2, M=8$};

\addplot [color=green!60!black,thick,solid,mark=o,mark options={solid}]
  table[row sep=crcr]{1	0.975950988132883\\
2	0.967189801361378\\
3	0.921456227236016\\
4	0.747410208727514\\
5	0.68956286572574\\
6	0.674833425286154\\
7	0.67030850215109\\
8	0.666216692738701\\
9	0.661212931404829\\
10	0.657472675051827\\
11	0.655502512073719\\
12	0.656118123870094\\
13	0.654557907539559\\
14	0.653165942868934\\
15	0.652060859603052\\
16	0.652564626167538\\
17	0.651540199083785\\
18	0.6517891169437\\
19	0.651677337762022\\
20	0.650852789990497\\
21	0.650261448297362\\
22	0.650570031961349\\
23	0.650464858339561\\
24	0.65053531316894\\
};
\addlegendentry{$N=3, M={16}$};

\addplot [color=orange!80!black,thick,solid,mark=square,mark options={solid}]
  table[row sep=crcr]{1	0.999999999999999\\
2	0.891705769217509\\
3	0.650163681339415\\
4	0.522359136216891\\
5	0.50161956239136\\
6	0.484498835915948\\
7	0.477632990910393\\
8	0.451711749969516\\
9	0.463537817029118\\
10	0.457052204728253\\
11	0.4587443349354\\
12	0.448962556864313\\
13	0.455398638922893\\
14	0.453624142755014\\
15	0.452381737637979\\
16	0.451038907509588\\
17	0.451479992079157\\
18	0.451102256801352\\
19	0.450713569387325\\
20	0.450399345802028\\
21	0.449789173565718\\
22	0.449557864095017\\
23	0.449526154421912\\
24	0.448985820391196\\
};
\addlegendentry{$N=4, M={16}$};

\addplot [color=orange!80!black,thick,solid,mark=triangle,mark options={solid}]
  table[row sep=crcr]{1	0.999421162339255\\
2	0.964391875539023\\
3	0.922779880297039\\
4	0.812223849998751\\
5	0.759373688941551\\
6	0.732371874642376\\
7	0.714964554231613\\
8	0.709216764678815\\
9	0.704903707375441\\
10	0.702169911612881\\
11	0.699786002802456\\
12	0.696293707264188\\
13	0.69487097422073\\
14	0.694361147093817\\
15	0.693414907342251\\
16	0.692637115592726\\
17	0.691865740435419\\
18	0.691609778798469\\
19	0.691231091681568\\
20	0.690713662933057\\
21	0.690073111462391\\
22	0.689977012761998\\
23	0.689726190029353\\
24	0.689360656936769\\
};
\addlegendentry{$N=4, M={64}$};

\addplot [color=blue,thick,dotted,forget plot]
  table[row sep=crcr]{1	0.79497665665095\\
2	0.79497665665095\\
3	0.79497665665095\\
4	0.79497665665095\\
5	0.79497665665095\\
6	0.79497665665095\\
7	0.79497665665095\\
8	0.79497665665095\\
9	0.79497665665095\\
10	0.79497665665095\\
11	0.79497665665095\\
12	0.79497665665095\\
13	0.79497665665095\\
14	0.79497665665095\\
15	0.79497665665095\\
16	0.79497665665095\\
17	0.79497665665095\\
18	0.79497665665095\\
19	0.79497665665095\\
20	0.79497665665095\\
21	0.79497665665095\\
22	0.79497665665095\\
23	0.79497665665095\\
24	0.79497665665095\\
};
\addplot [color=green!60!black,thick,dotted,forget plot]
  table[row sep=crcr]{1	0.649094323885851\\
2	0.649094323885851\\
3	0.649094323885851\\
4	0.649094323885851\\
5	0.649094323885851\\
6	0.649094323885851\\
7	0.649094323885851\\
8	0.649094323885851\\
9	0.649094323885851\\
10	0.649094323885851\\
11	0.649094323885851\\
12	0.649094323885851\\
13	0.649094323885851\\
14	0.649094323885851\\
15	0.649094323885851\\
16	0.649094323885851\\
17	0.649094323885851\\
18	0.649094323885851\\
19	0.649094323885851\\
20	0.649094323885851\\
21	0.649094323885851\\
22	0.649094323885851\\
23	0.649094323885851\\
24	0.649094323885851\\
};
\addplot [color=orange!80!black,thick,dotted,forget plot]
  table[row sep=crcr]{1	0.447213742853587\\
2	0.447213742853587\\
3	0.447213742853587\\
4	0.447213742853587\\
5	0.447213742853587\\
6	0.447213742853587\\
7	0.447213742853587\\
8	0.447213742853587\\
9	0.447213742853587\\
10	0.447213742853587\\
11	0.447213742853587\\
12	0.447213742853587\\
13	0.447213742853587\\
14	0.447213742853587\\
15	0.447213742853587\\
16	0.447213742853587\\
17	0.447213742853587\\
18	0.447213742853587\\
19	0.447213742853587\\
20	0.447213742853587\\
21	0.447213742853587\\
22	0.447213742853587\\
23	0.447213742853587\\
24	0.447213742853587\\
};
\addplot [color=orange!80!black,thick,dotted,forget plot]
  table[row sep=crcr]{1	0.687560042257491\\
2	0.687560042257491\\
3	0.687560042257491\\
4	0.687560042257491\\
5	0.687560042257491\\
6	0.687560042257491\\
7	0.687560042257491\\
8	0.687560042257491\\
9	0.687560042257491\\
10	0.687560042257491\\
11	0.687560042257491\\
12	0.687560042257491\\
13	0.687560042257491\\
14	0.687560042257491\\
15	0.687560042257491\\
16	0.687560042257491\\
17	0.687560042257491\\
18	0.687560042257491\\
19	0.687560042257491\\
20	0.687560042257491\\
21	0.687560042257491\\
22	0.687560042257491\\
23	0.687560042257491\\
24	0.687560042257491\\
};
\end{axis}
\end{tikzpicture}%
\end{small}
\caption{\label{fig:Approx}Obtained coherence over the number of approximation points $K$. Dotted lines indicate the results of an elaborate numerical integration.}
\end{figure}
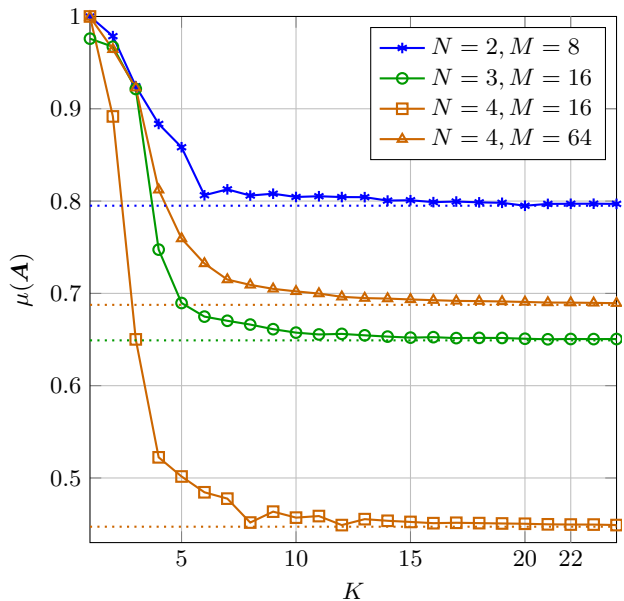

As mentioned before in Section~\ref{sec:Approx}, the potentially time consuming numerical computation of the integral in~\eqref{eq:Integral} can be relaxed to the sum given in~\eqref{eq:SumApprox}.
In order to evaluate the influence of numerical integration, we increased the number $K$ of points in the approximation for the constellations given in Table~\ref{tab:CoherenceTabXiLove}, and we plotted the best coherence out of ten runs in Fig.~\ref{fig:Approx}.
The coherence converges with increasing $K$.
Note that the convergence is not monotonic (cf. see $K=8$ in Fig.~\ref{fig:Approx} for $N=4, M=64$).
This is caused by the point distribution on the unit circle implied by $K$.
The gaps between the $K$ points favor vector sets with increased or reduced coherence.
The result of an optimization with the QAG adaptive integration from the GSL is indicated by dotted lines for reference.
For $K=22$, we examined the gained performance in more detail by Table~\ref{tab:Approx}.
\begin{table}[!t]
\renewcommand{\arraystretch}{1.3}
\centering
\caption{\label{tab:Approx}Evaluation of numerical integration with respect to running time and obtained coherence}
\begin{tabular}{r|r|c|c|c|c}
\hlx{hv[1,3,5]}
\multicolumn{2}{c|}{}&\multicolumn{2}{c|}{QAG adapt. integration}&\multicolumn{2}{c}{Summation}\\\hlx{v[1,3,5]hv}
$N$&$M$&Coherence&time $[\unit{s}]$&Coherence&time $[\unit{s}]$\\\hlx{vhv}
$2$&	$8$&	$0.7950$&	$699.03$&	$0.7971$&	$419.47$\\\hlx{vhv}
$3$&	$16$&	$0.6491$&	$2903.86$&	$0.6506$&	$1132.97$\\\hlx{vhv}
$4$&	$16$&	$0.4472$&	$27.26$&	$0.4496$&	$89.28$\\\hlx{vhv}
$4$&	$64$&	$0.6869$&	$64720.95$&	$0.6892$&	$3627.88$\\\hlx{vh}
\end{tabular}
\end{table}
Therein, the obtained coherence and corresponding running time in seconds is given.
As noted before, the elaborate integration reaches the Welch bound for $N=4, M=16$, whilst the simple summation over $K=22$ points does not and needs even more time.
However, in the other cases, summation is generally faster, especially for large values of $N$ and $M$.
The cause of this is again the point distribution on the unit circle, since the gaps between the $K$ points mislead the optimization especially in cases where the actual solution is easily found (there are no error bars observable in Fig.~\ref{fig:Dim4} for $M=16$).
The coherence values obtained by the simple summation are generally slightly worse.
With exception of the Welch bound achieving case $N=4$ and $M=16$, the coherence is still better than the results of other numerical approaches given in Table~\ref{tab:CoherenceTabXiLove}.
Interestingly, the approximation is also slower for the case of $N=4$ and $M=16$ which indicates that the approximation by $K=22$ points hindered the algorithm to converge properly in this special case.
As consequence, the choice of the numerical integration algorithm depends on the needed level of accuracy, where faster variants might be a suitable alternative, especially for demanding cases where $M\gg N^2$.
However, since the search for BCASCs is typically performed off line, accuracy is usually preferred over running time. 

\section{Conclusion}
\label{sec:Conclusion}
Within this contribution, we proposed a new approach to optimize the coherence of complex vector sets. 
Due to their tight relation, the presented results are also valuable in the fields of Grassmannian line packing and frame theory.
The results of our presented algorithm show for a wide range of vector sets improved coherence values compared to other algorithms.
Typically vector sets are searched off line and running time is not a major concern.
However, the potential drawback of increased computational effort can be countered by utilizing faster numerical integration algorithms which yield only slightly worse coherence results.

\section*{Acknowledgments}
The authors would like to thank Dejan Lazich for the fruitful discussions and the continuous support as well as the anonymous reviewers for their insightful and constructive comments which helped to improve this paper.

\appendix
\label{app:GlobMin}
For the complex-valued case, the Lagrange function stays the same [cf.~\eqref{eq:LagraneFunction}], and is still real-valued, as well as the Lagrange multipliers~$\boldsymbol{\lambda}$ and the constraint functions $\left\{\| \boldsymbol{s}_m\|^2 - 1=0\right\}_{m=1}^M$. 
However, due to the complex codewords, the necessary conditions for a global minimum of the potential function (\ref{eq:genpotfunc}) are now
\begin{align}
\nonumber	 \frac{\partial g(C_s(N,M),\boldsymbol{\lambda})}{\partial s_{mn}^{\operatorname{R}}} &= 0, \quad 
	 \frac{\partial g(C_s(N,M),\boldsymbol{\lambda})}{\partial s_{mn}^{\operatorname{I}}} = 0, \\
	 \frac{\partial g(C_s(N,M),\boldsymbol{\lambda})}{\partial \lambda_m} &= 0,
\end{align}
with $m = 1,\hdots,M$ and $n = 1,\hdots,N$.
For easier expressions, we introduce the following abbreviations: $\delta_{ml}=\| \boldsymbol{s}_m - \boldsymbol{s}_l\|$ and $f(x)=x^{-(\nu-2)}$.
Thus, the generalized potential function can be expressed by $g(C_s(N,M))=\sum_{m=1}^{M}\sum_{l<m}f(\delta_{ml})$, which leads with ${\partial\delta_{ml}}/{\partial s_{mn}^{\operatorname{R}}}={(s_{mn}^{\operatorname{R}}-s_{ln}^{\operatorname{R}})}/{\delta_{ml}}$ to
\begin{align}
\nonumber\frac{\partial g(C_s(N,M),\boldsymbol{\lambda})}{\partial s_{mn}^{\operatorname{R}}} =& 0\\
\nonumber=&\frac{\partial \lambda_m \left( {s_{mn}^{\operatorname{R}^2}} + {s_{mn}^{\operatorname{I}^2}} - 1\right)}{\partial s_{mn}^{\operatorname{R}}} + \sum_{l\neq m}\frac{\partial f(\delta_{ml})}{\partial s_{mn}^{\operatorname{R}}}\\
\nonumber=& 2s_{mn}^{\operatorname{R}}\lambda_m + \sum_{l\neq m}\frac{\partial f(\delta_{ml})}{\partial\delta_{ml}}\cdot \frac{s_{mn}^{\operatorname{R}}-s_{ln}^{\operatorname{R}}}{\delta_{ml}}\\
\nonumber=& 2s_{mn}^{\operatorname{R}}\lambda_m - (\nu-2)\sum_{l\neq m}\frac{s_{mn}^{\operatorname{R}}-s_{ln}^{\operatorname{R}}}{\delta_{ml}^{\nu}}\\
\Rightarrow s_{mn}^{\operatorname{R}}=&\frac{(\nu-2)\sum_{l\neq m}\frac{s_{mn}^{\operatorname{R}}-s_{ln}^{\operatorname{R}}}{\delta_{ml}^{\nu}}}{2\lambda_m}.
\end{align}
The calculation of $s_{mn}^{\operatorname{I}}$ is equivalent with the imaginary instead of the real part.
Therefore, the remaining necessary condition ${\partial g(C_s(N,M),\boldsymbol{\lambda})}/{\partial \lambda_m} = 0$ gives
\begin{align}
\nonumber 1&=\sum_{k=1}^{N}{s_{mk}^{\operatorname{R}}}^2+{s_{mk}^{\operatorname{I}}}^2\\
\nonumber&=\sum_{k=1}^{N}{\left( \frac{\sum_{l\neq m}\frac{(\nu-2)(s_{mk}^{\operatorname{R}}-s_{lk}^{\operatorname{R}})}{\delta_{ml}^{\nu}}}{2\lambda_m} \right)}^2\\
\nonumber&\phantom{\sum_{k=1}^{N}}+{\left(\frac{\sum_{l\neq m}\frac{(\nu-2)(s_{mk}^{\operatorname{I}}-s_{lk}^{\operatorname{I}})}{\delta_{ml}^{\nu}}}{2\lambda_m}\right)}^2\\
\nonumber&=\frac{(\nu-2)^2}{4\lambda^2_m}\sum_{k=1}^{N}{\left( \sum_{l\neq m}\frac{s_{mk}^{\operatorname{R}}-s_{lk}^{\operatorname{R}}}{\delta_{ml}^{\nu}} \right)}^2\\
\nonumber&\phantom{\sum_{k=1}^{N}}+{\left(\sum_{l\neq m}\frac{s_{mk}^{\operatorname{I}}-s_{lk}^{\operatorname{I}}}{\delta_{ml}^{\nu}}\right)}^2\\
\nonumber&=\frac{(\nu-2)^2}{4\lambda^2_m}\left\| \sum_{l\neq m}\frac{\boldsymbol{s}_{m}-\boldsymbol{s}_{l}}{\delta_{ml}^{\nu}} \right\|^2\\
\Rightarrow2\lambda_m&=(\nu-2) \left\| \sum_{l\neq m}\frac{\boldsymbol{s}_{m}-\boldsymbol{s}_{l}}{\delta_{ml}^{\nu}} \right\|.
\end{align}
Thus, we can give the equilibrium of a codeword element
\begin{align}
\nonumber s_{mk}
&=s_{mk}^{\operatorname{R}}+\mathrm{i}\cdot s_{mk}^{\operatorname{I}}\\
\nonumber&=\frac{(\nu-2)\sum_{l\neq m}\frac{s_{mk}^{\operatorname{R}}-s_{lk}^{\operatorname{R}}}{\delta_{ml}^{\nu}}
+\mathrm{i}\cdot\frac{s_{mk}^{\operatorname{I}}-s_{lk}^{\operatorname{I}}}{\delta_{ml}^{\nu}}}{2\lambda_m}\\
&=\frac{\sum_{l\neq m}\frac{s_{mk}-s_{lk}}{\delta_{ml}^{\nu}}}{\left\| \sum_{l\neq m}\frac{\boldsymbol{s}_{m}-\boldsymbol{s}_{l}}{\delta_{ml}^{\nu}} \right\|}
\end{align}
which can be combined to the equilibrium of~\eqref{eq:erdw}.
Consequently, the global minimum of the generalized potential function $g(C_s(N,M))$ can be expressed by the equilibrium~\eqref{eq:erdw} also in the case of complex vector spaces.

\bibliographystyle{IEEEtran}

\end{document}